\documentclass[aps,prl,reprint,superscriptaddress,amsmath,amssymb]{revtex4-2}

\usepackage{graphicx}
\usepackage{bm}
\usepackage[colorlinks,linkcolor=blue,citecolor=blue,urlcolor=blue]{hyperref}
\usepackage{braket}
\usepackage{verbatim}

\newcommand*{\ShortSecTitle}[1]{\textit{\color{blue}#1}.---}

\begin{document}

\title{Emergent Dynamical Scale Invariance in Long-Range Quantum Spin Chains}

\author{Xuhui Yao}
\affiliation{Department of Physics, Fudan University, Shanghai 200433, China}

\author{Ning Sun}
\email{ningsun.atom@gmail.com}
\affiliation{Department of Physics, Fudan University, Shanghai 200433, China}
\affiliation{Department of Physics, The University of Hong Kong, Hong Kong, China}

\author{Lei Feng}
\email{leifeng@fudan.edu.cn}
\affiliation{Department of Physics, Fudan University, Shanghai 200433, China}
\affiliation{Institute of Micronano Electronic Devices and Quantum Computing and Shanghai Key Laboratory of Meta-surface Optoelectronic Fine Control, Fudan University, Shanghai 200438, China}
\affiliation{Hefei National Laboratory, Hefei 230088, China}
\date{\today}

\begin{abstract}

Universality is central to physics, exemplified by few-body phenomena in ultracold atomic gases. Quantum simulation opens a new setting in long-range quantum spin systems, where tunable magnon dispersions support distinct universality classes. Universal bound states and correlation relations are established, but these classes' manifestation in real-time dynamics remains open. We demonstrate emergent dynamical scale invariance in two-magnon dynamics of long-range quantum spin chains with \(1/r^\alpha\) exchange couplings. The mechanism is a strongly interacting two-body resonance enabled by nonquadratic one-dimensional magnon dispersion. An analytical two-body solution shows that appropriately rescaled initial wave packets evolve into density profiles related by continuous space-time scaling in the low-energy limit. Both translation-invariant and impurity models exhibit this behavior over broad resonant regimes, \(3/2<\alpha<2\) and \(2<\alpha<3\), respectively. Numerical simulations verify these predictions and show profile collapse in both models. We propose a trapped-ion protocol to realize these dynamics and probe few-body universality beyond quadratic dispersion. This nonequilibrium perspective goes beyond traditional spectroscopy, offering new opportunities for quantum simulators.

\end{abstract}

\maketitle

\begin{figure}[t]
    \centering
    \includegraphics[width=\columnwidth]{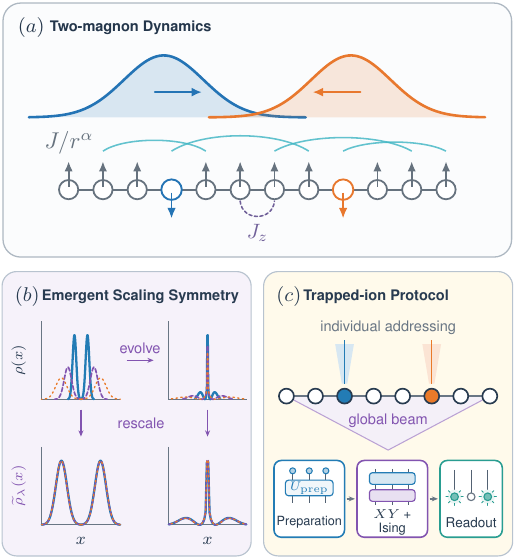}
    \caption{Schematic of the emergent dynamical scaling symmetry in long-range spin chain. (a) Counter-propagating magnon packets in a spin chain with long-range \(XY\) exchange \(J/r^\alpha\) and nearest-neighbor $ZZ$ interaction \(J_z\). (b) In strongly interacting resonant regime, the two-body dynamics exhibit emergent scaling symmetry. (c) We propose a concrete trapped-ion protocol for experimental implementation.}
    \label{fig:1}
\end{figure}

Universality is a central concept in modern physics: systems with different microscopic constituents can exhibit identical behavior at appropriate scales~\cite{Wilson1975,Stanley1999}. In few-body physics, this means that low-energy scattering and binding are governed by a small number of parameters, largely independent of microscopic details~\cite{BraatenHammer2006,NaidonEndo2017}. The Efimov effect is a paradigmatic example, in which resonant two-body interactions produce an infinite sequence of three-body bound states with discrete scale invariance~\cite{Efimov1970,Bedaque1999}. Ultracold atomic gases, enabled by tunable Feshbach resonances~\cite{Chin2010}, have provided a primary platform for studying these phenomena~\cite{Kraemer2006,Zaccanti2009,Pollack2009,Huang2014,Pires2014,Tung2014}. However, most universal few-body physics has focused on nonrelativistic particles with quadratic dispersion and short-range interactions.

Quantum simulators now offer controlled settings to explore universality beyond this conventional regime. In particular, trapped-ion systems realize quantum spin chains with tunable long-range interactions,
$J_{mn}\simeq J_0|m-n|^{-\alpha}$,
together with local preparation and detection of spin excitations~\cite{Porras2004,Kim2009,Britton2012,Monroe2021,Jurcevic2014,Richerme2014,Senko2014}. Recent advances have extended this capability to two-dimensional arrays of hundreds of ions with site-resolved readout~\cite{Guo2024}. Although these platforms have primarily been used to study many-body phenomena~\cite{Kim2010,Islam2013,Feng2023,Smith2016,Zhang2017Dynamics,Morong2021,Brydges2019,Joshi2022}, their control also enables the isolation of a few magnons and the investigation of their interactions~\cite{Fukuhara2013,Kranzl2023}.

Our recent works showed that long-range quantum spin systems support distinct universal few-body regimes~\cite{Sun2026Efimov,Sun2026Impurity,Sun2026Contact}. Their low-energy magnon dispersion,
$\epsilon_k\simeq u|k|^z$,
has a tunable exponent $z$, which modifies the conditions for resonant scattering and scale invariance~\cite{Wang2022,Sun2026Efimov}. Consequently, Efimov physics can arise in a one-dimensional spin chain~\cite{Sun2026Efimov}, while a local impurity supports Efimov and semi-super-Efimov spectra~\cite{Sun2026Impurity}. Universal contact relations further connect spin correlations with the high-frequency dynamical structure factor~\cite{Sun2026Contact}. These results establish universal spectra, scattering properties, and correlations, but leave open how such resonant universality classes appear in real-time dynamics.

Recent experiments have demonstrated Efimovian and super-Efimovian expansion~\cite{Deng2016,Shi2017,Deng2018}, while interaction quenches reveal universal dynamics and the formation of few-body correlations~\cite{Makotyn2014,Eigen2017,Eigen2018,Sykes2014,Corson2015,Corson2016,Colussi2018,Sun2020}. 
In quantum spin systems, experiments have demonstrated
the controlled propagation and binding of spin
excitations~\cite{Jurcevic2014,Fukuhara2013}.
In particular, magnon bound states have been observed
in a trapped-ion realization of the long-range anisotropic
Heisenberg model~\cite{Kranzl2023}.
These developments motivate the question: can resonantly interacting magnons in long-range quantum spin systems exhibit universal dynamical scaling governed by a nonquadratic dispersion?

In this Letter, we show emergent dynamical scale invariance in the two-magnon dynamics of long-range quantum spin chains, which originates from the strongly interacting two-body resonance enabled by the tunable magnon dispersion in one dimension.
In the low-energy limit, appropriately rescaled initial wave packets evolve into density profiles related by the continuous transformation $x\to \lambda x$, $t\to \lambda^z t$.
We study both translation-invariant model and impurity model, showing that the dynamical scale invariance emerges in both systems, where the corresponding resonant regimes extend over a wide range of $3/2<\alpha<2$ and $2<\alpha<3$, respectively.
Analytical scaling arguments and resonance conditions of the lattice model are also verified by numerical simulations
by collapsing density profiles via proper scaling.
Moreover, we propose a concrete trapped-ion protocol for realizing the two-magnon quench dynamics in experiment.
Our results establish a dynamical manifestation of few-body universality beyond the conventional quadratic-dispersion setting and provide a route to probing it in quantum simulators.

\ShortSecTitle{Model}
We consider the one-dimensional 
spin chain described by the Hamiltonian $H=H_{XY}+H_I$, where $H_{XY}$ is the isotropic long-range spin exchange interaction in $xy$-plane given by 
\begin{equation}\label{eqn:hxy}
H_{XY}=-\sum_{i<j}J_{ij}
\left(\sigma_i^x\sigma_j^x+\sigma_i^y\sigma_j^y\right),
\end{equation}
with $J_{ij}=J(|i-j|)^{-\alpha}$.
$H_I$ is the homogeneous or impurity nearest-neighbor interaction along $z$-direction, $H_I^{\mathrm{hom}}$ or $H_I^{\mathrm{imp}}$, respectively,
\begin{equation}\label{eqn:hz}
\begin{aligned}
&H_I^{\mathrm{hom}}=-J_z\sum_i\sigma_i^z\sigma_{i+1}^z,\\
&H_I^{\mathrm{imp}}
= -J_z\sigma_{m}^z\sigma_{m+1}^z + \Delta(\sigma_m^z+\sigma_{m+1}^z). 
\end{aligned}
\end{equation}
where $\sigma_i^{n}$ with $n\in\{x,y,z\}$, and $\Delta=J_z$. The local detuning $\Delta$ effectively eliminates the one-body local potential brought by the local $ZZ$ term. Thus the impurity interaction is equivalent to $-4J_zP_m^\downarrow P_{m+1}^\downarrow + const.$ with the projection operator $P=\ket{\downarrow}\bra{\downarrow}$.
Without loss of generality, we place the impurity, the $(m,m+1)$-bond, at the center of the chain.
Because both the Hamiltonian in Eqs.~\eqref{eqn:hxy} and ~\eqref{eqn:hz} possess rotational symmetry along the $z$ direction, 
the total magnetization $S_z=\sum_i\sigma_i^z$ conserves in both models.

To gain better insights on the low energy excitations, we introduce a quasiparticle description by taking the fully polarized state $|g\rangle=\otimes_i\ket{\uparrow}_i$ as the reference state (vacuum) and a single spin flip $\sigma^-|g\rangle$ as a magnon excitation (particle). 
We have previously derived the single magnon dispersion in momentum space, which is $\epsilon_k=-\sum_{l}^\infty4J\cos(kl)/(l)^\alpha$ ~\cite{Sun2026Efimov,Sun2026Impurity}. 
When $\alpha>3$, this result reduces to the normal quadratic dispersion because the leading order in the low-energy limit ($|k|\ll1$) is $\epsilon_k\approx\epsilon_0 + u|k|^z$ with $z=\min\{\alpha-1,2\}$.
In contrast, when $1<\alpha<3$, the single particle dispersion is non-quadratic, which sets the stage for new physics. 
In addition, the total $S_z$ conservation implies total magnon number conservation and we can further restrict our study within a fixed-magnon-number subspace.
We extensively study the universal few-body(magnon) physics of the distinct non-quadratic universality class~\cite{Sun2026Efimov,Sun2026Impurity,Sun2026Contact,huo2026} on such a long-range spin system.
In this work, we focus on the two-magnon sector. 

\ShortSecTitle{Two-body resonance}
Two-magnon resonance occurs when a bound state emerges at zero energy when we tune the interaction $J_z$ to the critical point $J_z^*$, which is analogous to the Feshbach resonance in cold atoms. At the resonance, we can treat the spin system as a dilute magnon gas in the unitary limit. More precisely in the homogeneous model, the low-energy two-magnon scattering $T$-matrix takes the form $T(E)^{-1}=P^{-1}-B_1(z)(-E_+)^{(1-z)/z}$ at zero total momentum, with $E_+=E+i0^+$, $B_1(z)$ a $z$-dependent constant, and $P$ the generalized scattering length characterizing the low-energy collisions~\cite{Sun2026Efimov,Sun2026Contact}. In the interval of $1/2<z<1$ (i.e., $3/2<\alpha<2$), tuning $J_z
$ to $J_z^*$ drives $P$ to infinity~\cite{Sun2026Efimov} and thus enables two magnon resonance. 
In the impurity model, local interaction term of $\sigma_m^z\sigma_{m+1}^z$ can also introduce one-body resonances. We use specially designed local terms of Eq.~\eqref{eqn:hz} together with Eq.~\eqref{eqn:hxy} to eliminate such effect and only enable magnons scattering via the impurity-mediated two-body interaction. The corresponding $T$-matrix is $T(E)^{-1}=G^{-1}-B_2(z)(-E_+)^{(2-z)/z}$, with the generalized scattering length $G$ and constant $B_2(z)$~\cite{Sun2026Impurity}. 
Now the two-magnon resonance occurs in the window $1<z<2$ (i.e., $2<\alpha<3$). Tuning $J_z$ to the critical value $J_z^*$ similarly drives $G$ to infinity at the two magnon resonance. The details are derived in the End Matter.

\begin{figure}[t]
\centering
\includegraphics[width=\columnwidth]{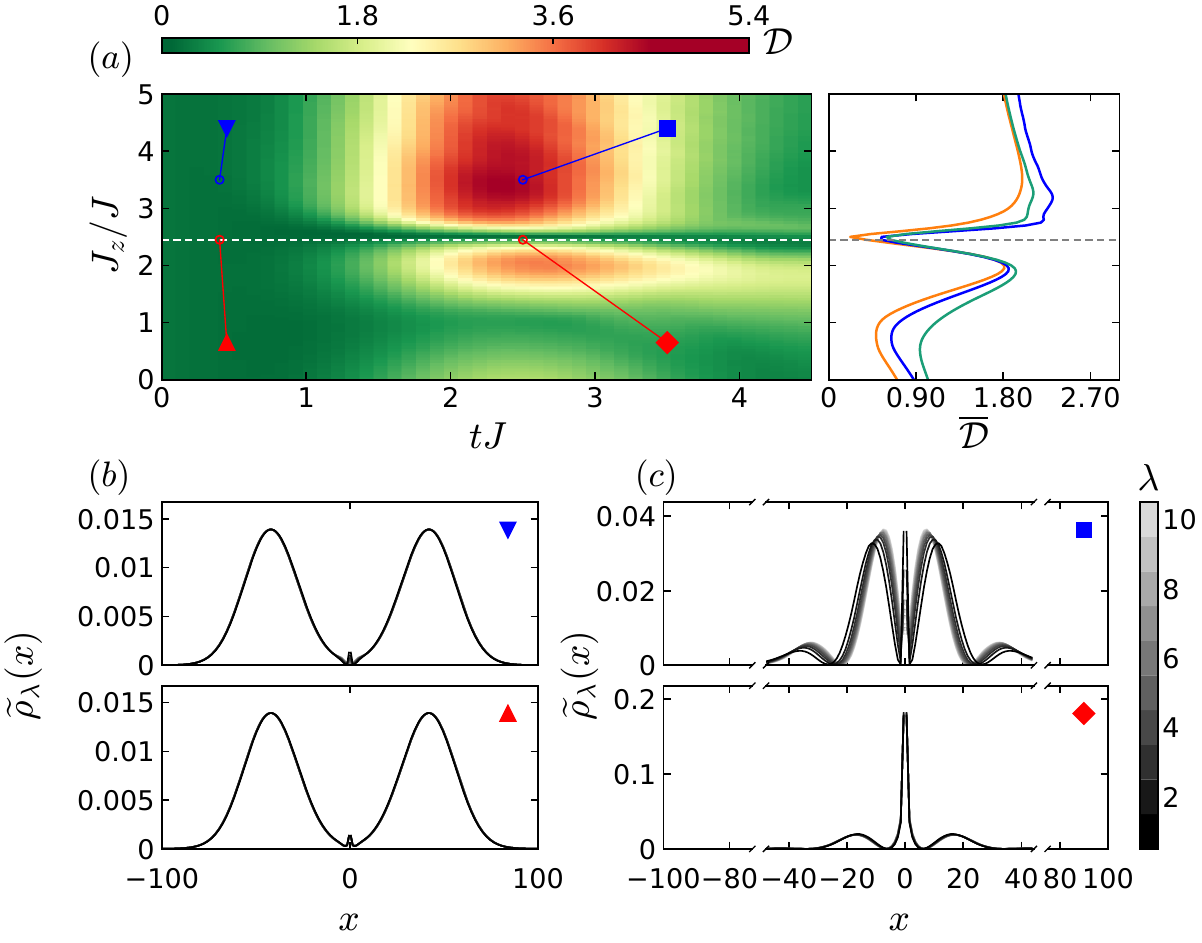}
\caption{Two-magnon dynamics and scaling distance $\mathcal{D}$ in the homogeneous spin chain with $\alpha=1.75$ and total momentum $Q=0$. 
The scaling distance is calculated using 9 scaled initial states with $\lambda=2\sim10$ incrementally, which are evolved using relative-coordinate cutoffs $L=1000$.
(a) Left: The heat plot of scaling distance for Gaussian initial wave packets at different $J_z/J$ and evolution time $t$.  Red(blue) markers indicate the (off-)resonance regime. Right: the time-averaged distance $\overline{\mathcal{D}}$ for Gaussian (orange), sinc (blue), and super-Gaussian (green) initial wave packets.
Dashed lines in (a) mark the analytical value of the two-magnon resonance $J_z^*/J=2.446$. 
(b),(c): The rescaled density profile $\widetilde{\rho}_{\lambda}(x)$ of different scaling $\lambda$ at evolution time $tJ=0.4$ (b) and $tJ=2.5$ (c). 
Upper panels of (b),(c): off-resonant regime $J_z/J=3.5$ (blue triangle and blue square). Lower panels of (b),(c): strongly interacting resonance regime $J_z/J=2.45$ (red triangle and red diamond).
}
\label{fig:2}
\end{figure}

\begin{figure}[t]
\centering
\includegraphics[width=\columnwidth]{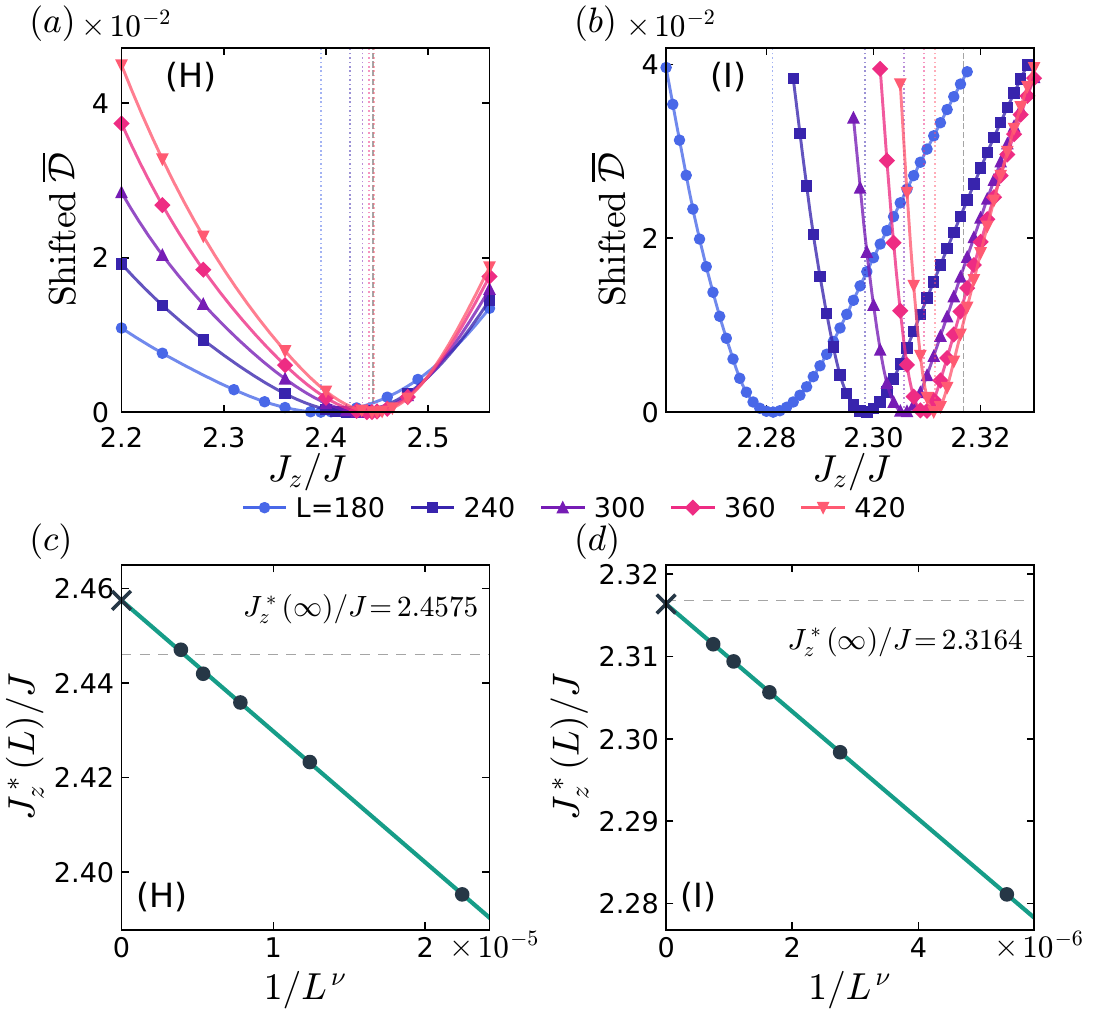}
\caption{
Finite-size dependence of the SD minima in open chains,
using rescaled one-body density.
(a),(b) Shifted $\overline{\mathcal{D}}$ versus $J_z/J$
for the homogeneous ($\alpha=1.75$) and impurity
($\alpha=2.5$) models, respectively.
Each curve is shifted by subtracting its interpolated minimum.
Colored dotted lines mark the interpolated minima $J_z^*(L)$.
(c),(d) $J_z^*(L)/J$ versus $L^{-\nu}$ with linear fits.
Crosses mark the extrapolated SD-minimum positions:
$J_z^*(\infty)/J=2.4575$ with $\nu=2.061$ (homogeneous)
and $2.3164$ with $\nu=2.336$ (impurity).
Gray dashed lines indicate the analytical zero-energy
resonances at $J_z^*/J\simeq2.446$ and $2.317$, respectively.
}
\label{fig:3}
\end{figure}

\ShortSecTitle{Two-magnon dynamics and emergent scaling symmetry}
In general, we can fully characterize the two-magnon low-energy physics by one scattering parameter $P$ or $G$, and all physical quantities can in principle be written as functions of $P$, or $G$, together with spacetime $\{x,t\}$. 
However, tuning $J_z$ to the critical value $J_z^*$, namely to the two magnon resonance, drives $P$ or $G$ to infinity.
Thus, it leaves no characteristic length scale to describe the system, hence imply a scale invariance. As a concrete demonstration, 
we analytically solve
the two-magnon quench dynamics in the zero total quasimomentum $Q=0$ sector of the
homogeneous model (See End Matter). 
The solution at the two-magnon resonance in the low-energy continuum limit rigorously obeys
\begin{equation}
\psi(\lambda x,\lambda^z t)
=\lambda^{-1/2}\psi(x,t),
\label{eq:dynamical_scaling}
\end{equation}
with $\psi(x,t)$ the two-magnon wavefunction and $x$ the separation between the two magnons.
Consequently,
$\int\rho(\lambda x,\lambda^z t)\,d(\lambda x)
=\int\rho(x,t)\,dx = 1$,
where $\rho(x,t)=|\psi(x,t)|^2$ is the probability density.

To measure the scale invariance in numerics, we introduce the scaling distance (SD), denoted by $\mathcal{D}$, to quantify the level of the distance between any given state $|\psi\rangle$ and its rescaled counterparts.
Hence smaller SD means more likeliness between states with the rescaled thereof. By evaluating the SD along the dynamics, $|\psi(t)\rangle=e^{-iHt}|\psi_0\rangle$, we can assess if there is and how well is the scale invariance. 
Specifically, it is introduced as
\begin{equation}
\begin{aligned}
    \mathcal D(t)&=
    \sum_{\lambda}
    \int dx
    \left|\rho(x,t)-\widetilde{\rho}_{\lambda}(x,t)
    \right|,
    \label{eq:scaling_distance}
\end{aligned}
\end{equation}
with $\widetilde{\rho}_{\lambda}(x,t)=\lambda\rho(\lambda x,\lambda^zt)$.
For lattice simulations, probabilities are coarse-grained onto a grid, and the integral is replaced by a sum over the resulting bin probabilities.
This gives us a robust measure of the likeliness between states since we sample through multiple scaling factors $\lambda$.

We first numerically show the scaling invariance for the homogeneous model in the $Q=0$ sector, with $\alpha=1.75$ without loss of generality.
We prepare a low-energy Gaussian wave packet in the relative
coordinate,
\begin{equation}
\psi(x,0)=\mathcal{N}
\exp\left[-\frac{(|x|-x_0)^2}{2\sigma_0^2}
-ik_0|x|\right],
\label{eqn:initial}
\end{equation}
where $\mathcal{N}$ ensures normalization, and $x=0$ is excluded
by the hard-core constraint. 
Numerically, we choose $x_0=50$, $\sigma_0=20$, and $k_0=0.1$
to probe the low-energy regime, and sample $\lambda$ over 9 different values
within $[2,10]$ to evaluate the SD $\mathcal{D}$. We truncate the relative coordinate at
$|x|\le L$, with $L=1000$,
and normalize each state on its discrete grid. The numerical results are shown in Fig.~\ref{fig:2}. 
 We calculate the two-magnon dynamics across different $J_z/J$. The SD shows consistent minima at the resonance at $J_z^*/J = 2.446$ marked by a white dashed line to guide the eye. 
The time-averaged SD $\overline{\mathcal{D}}=(1/T)\int_0^T\mathcal{D}(t)\,dt$
with $TJ=4.5$ shows a prominent dip and reinforces this conclusion.
We further present a more intuitive demonstration of the scaling invariance by directly showing the magnon density at $tJ=0.4$ and $tJ=2.5$ in Fig.~\ref{fig:2}(b) and (c).
The (lack of) invariance is clearly shown as the (non-) perfect collapse of the density profiles after scaling at the resonance case, which is particularly distinct after the magnon collision at later times.
This result is robust against choices of initial wavefunction. As examples, additional choice of the initial envelope of sinc
($\propto\sin[k_0(|x|-x_0)]/[k_0(|x|-x_0)]$) or fourth-order
super-Gaussian ($\propto e^{-[(|x|-x_0)/\sigma_0)]^8}$) results in very similar dynamics.

Next, we investigate full two-magnon states without translation invariance, which are more relevant to experimental platforms.
Let $\Psi(x_1,x_2,t)$ denote the two-magnon
wavefunction in the continuum limit. At the two-magnon
resonance, we expect the low-energy dynamics to obey
$\Psi(\lambda x_1,\lambda x_2,\lambda^z t)
=\lambda^{-1}\Psi(x_1,x_2,t)$, which induces the corresponding scaling of the normalized
one-body density,
$\rho^{(1)}(x,t)
=\langle\Psi(t)|P_\downarrow(x)|\Psi(t)\rangle/2$.
Thus we can apply the same SD definition to
$\rho^{(1)}$ to measure the scale invariance.
In lattice simulation on finite open chains, we evolve two-magnon states from scaled Gaussian wave packets using exact
diagonalization for the homogeneous ($\alpha=1.75$) and
impurity ($\alpha=2.5$) models (see SM~\cite{SM} for details).
For each chain length $L$, we evaluate the SD at $\lambda=1,1.5,2,2.5,3$.
Repeating this calculation for $L=180$--$420$, we find that
the time-averaged SD exhibits a distinct local minimum at
$J_z^*(L)$ in both models.
The minima shift toward larger $J_z/J$ with increasing $L$,
revealing the finite-size effect (Figs.~\ref{fig:3}(a) and (b)).

To estimate the resonance locations in the infinite-size limit,
we fit the SD minima using the empirical form
$J_z^*(L)=J_z^*(\infty)+aL^{-\nu}$,
where $J_z^*(\infty)$ is the extrapolated SD-minimum position,
$a$ is the correction amplitude, and $\nu$ is the exponent.
The fits yield $J_z^*(\infty)/J=2.4575$ with $\nu=2.061$
for the homogeneous model and
$J_z^*(\infty)/J=2.3164$ with $\nu=2.336$ for the impurity
model (Figs.~\ref{fig:3}(c) and (d)).
Both extrapolated results are close to the analytic zero-energy
resonances at $J_z^*/J\simeq2.446$~\cite{Sun2026Efimov} and
$2.317$ (see End Matter).

\begin{figure}[t]
    \centering
    \includegraphics[width=\columnwidth]{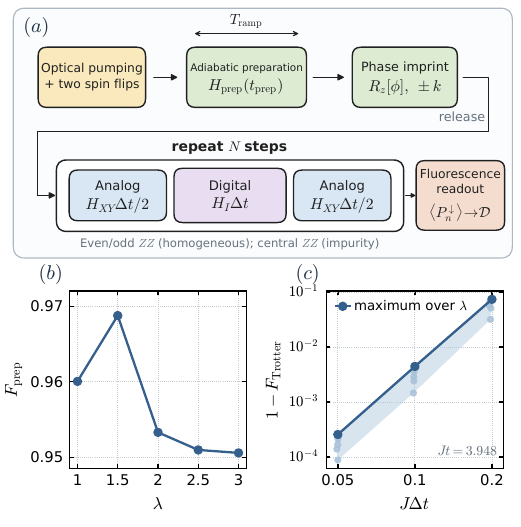}
    \caption{\label{fig:4}
    Hybrid analog--digital trapped-ion protocol for both models,
    with numerical benchmarks for the homogeneous model.
    (a) Adiabatic packet preparation, momentum imprinting by local $Z$ rotations, and Trotter evolution followed by fluorescence readout.
    (b) Preparation fidelity $F_{\mathrm{prep}}$ for five scales. Error bars show the absolute change upon reducing the preparation integration step from $0.5/J$ to $0.25/J$.
    (c) Final-state infidelity $1-F_{\mathrm{Trotter}}$ relative
    to exact evolution from the same prepared states, versus
    the Trotter step $\Delta t$.
    Faint markers show individual scales, shading spans their
    range, and the solid line gives their maximum.
    The evolution uses $\alpha=1.75$, $J_z/J=2.4$, and time $Jt\simeq3.948$.
    }
\end{figure}

\ShortSecTitle{Experimental protocol}
To test our theory in experiment, we propose a trapped-ion protocol for preparing scalable
two-magnon packets and evolving them with analog-digital
Trotter steps (Fig.~\ref{fig:4}(a)). 
For initial-state preparation, we propose an adiabatic scheme
using site-resolved AC-Stark shifts to create longitudinal
fields~\cite{Monroe2021,Morong2021}, which form a symmetric truncated-quadratic double-well
potential.
Starting with one spin flip at each minimum, we smoothly
turn on $H_{XY}$ while weakening the confinement.
The resulting ground state approximates two Gaussian
magnon packets, onto which local $Z$ rotations imprint
opposite momenta.

After the state preparation, we implement the evolution
of the two-magnon packets through the mixed analog and digital blocks in a Trotter sequence. The analog block utilizes a bichromatic spin-dependent force to generate long-range
M{\o}lmer--S{\o}rensen $XX$ interactions. A uniform strong field
$B_z\sum_n\sigma_n^z$ suppresses magnon-pair creation and annihilation in the rotating frame and gives a number-conserving exchange term of $XX+YY$~\cite{Wall2017,Joshi2022,Feng2023}.
The exponent $\alpha$ is tunable through the laser
detuning from the collective vibrational modes. The
digital block $H_I^{\rm hom}$ is implemented with
nearest-neighbor $ZZ$ gates in even- and odd-bond layers,
whereas $H_I^{\rm imp}$ only requires a central-pair $ZZ$ gate and
local $Z$ rotations. Such parallel entangling gates on disjoint ion pairs have been demonstrated~\cite{Figgatt2019,Zhu2023Parallel}.
After evolution,
site-resolved fluorescence readout provides the one-body
densities for computing the SD.

We numerically benchmark the preparation and evolution protocol in an open chain of $L=180$. The results show that with ramp durations up to $JT_{\rm ramp}=550$, the preparation
fidelity, $F_{\mathrm{prep}}
=|\langle\Psi(0)|
\Psi^{\mathrm{prep}}\rangle|^2$
exceeds $0.95$ across all five scales (Fig.~\ref{fig:4}(b)), where the two states denote the target
and prepared states, respectively. To isolate the Trotter error, we compare exact and Trotter
evolution from the same prepared state using
$F_{\mathrm{Trotter}}
=|\langle\Psi^{\mathrm{exact}}(\lambda^z t)|
\Psi^{\mathrm{Trotter}}(\lambda^z t)\rangle|^2$.
At Trotter step $J\Delta t\approx0.05$, the maximum final-state
infidelity $1-F_{\mathrm{Trotter}}$ across the five scales
is $2.56\times10^{-4}$ (Fig.~\ref{fig:4}(c)).
Note that these benchmarks assume ideal effective spin dynamics,
excluding finite-field corrections\cite{Wall2017}, residual spin--phonon
coupling, decoherence, and calibration
drift~\cite{Wang2012,Lewis2023}. A fully inclusive simulation is beyond the scope of this work.

\ShortSecTitle{Discussion}
In this Letter, we demonstrate emergent scaling symmetry of strongly interacting two-magnon dynamics in long-range quantum spin chains. At the two-body resonance, the dynamics of two magnon low-energy states exhibits a continuous spacetime scale invariance under $x\rightarrow \lambda x$, $t\rightarrow \lambda^zt$, with $z$ being the nonquadratic magnon dispersion exponent. 
These findings connect few-body universality, nonequilibrium quantum dynamics, and quantum magnetism beyond the conventional non-relativistic paradigm, extending their significance beyond spectral properties.
Besides, our work opens opportunities for future studies. First, implementing the proposed trapped-ion protocol would allow direct experimental tests of the emergent dynamical scaling symmetry. Second, studying quenches away from the resonance regime could reveal how an additional interaction scale controls the breakdown of universal dynamical scaling. Finally, extending the analysis to three-magnon dynamics raises an intriguing question of how Efimov scaling structures manifest in real-time evolution. These studies would further establish long-range quantum spin systems as a platform for studying universal few-body dynamics beyond the conventional quadratic-dispersion paradigm.

\textit{Acknowledgments.}
We thank Pengfei Zhang for helpful discussions. 
This project is supported by the Shanghai Municipal Science and Technology Major Project Grant No. 24DP2600100, Co-research Program under Grant No. 25LZ2601000, the Quantum Science and Technology–National Science and Technology Major Project under Grants No. 2023ZD0300900, No. 2025ZD0300100 and No. 2025ZD0300101, and Scientific Research Innovation Capability Support Project for Young Faculty under Grant No. ZYGXQNJSKYCXNLZCXM-I14.
N.S. also acknowledges supports from NSFC under Grant No. 12604404, HK GRF Grants No. 17306024, CRF Grants No. C7015-24GF and C4050-23GF, CRS\_HKU701/24, and a RGC Fellowship Award No. HKU RFS2223-7S03. 

\bibliography{ref}

\begin{thebibliography}{53}%
\makeatletter
\providecommand \@ifxundefined [1]{%
 \@ifx{#1\undefined}
}%
\providecommand \@ifnum [1]{%
 \ifnum #1\expandafter \@firstoftwo
 \else \expandafter \@secondoftwo
 \fi
}%
\providecommand \@ifx [1]{%
 \ifx #1\expandafter \@firstoftwo
 \else \expandafter \@secondoftwo
 \fi
}%
\providecommand \natexlab [1]{#1}%
\providecommand \enquote  [1]{``#1''}%
\providecommand \bibnamefont  [1]{#1}%
\providecommand \bibfnamefont [1]{#1}%
\providecommand \citenamefont [1]{#1}%
\providecommand \href@noop [0]{\@secondoftwo}%
\providecommand \href [0]{\begingroup \@sanitize@url \@href}%
\providecommand \@href[1]{\@@startlink{#1}\@@href}%
\providecommand \@@href[1]{\endgroup#1\@@endlink}%
\providecommand \@sanitize@url [0]{\catcode `\\12\catcode `\$12\catcode `\&12\catcode `\#12\catcode `\^12\catcode `\_12\catcode `\%12\relax}%
\providecommand \@@startlink[1]{}%
\providecommand \@@endlink[0]{}%
\providecommand \url  [0]{\begingroup\@sanitize@url \@url }%
\providecommand \@url [1]{\endgroup\@href {#1}{\urlprefix }}%
\providecommand \urlprefix  [0]{URL }%
\providecommand \Eprint [0]{\href }%
\providecommand \doibase [0]{https://doi.org/}%
\providecommand \selectlanguage [0]{\@gobble}%
\providecommand \bibinfo  [0]{\@secondoftwo}%
\providecommand \bibfield  [0]{\@secondoftwo}%
\providecommand \translation [1]{[#1]}%
\providecommand \BibitemOpen [0]{}%
\providecommand \bibitemStop [0]{}%
\providecommand \bibitemNoStop [0]{.\EOS\space}%
\providecommand \EOS [0]{\spacefactor3000\relax}%
\providecommand \BibitemShut  [1]{\csname bibitem#1\endcsname}%
\let\auto@bib@innerbib\@empty
\bibitem [{\citenamefont {Wilson}(1975)}]{Wilson1975}%
  \BibitemOpen
  \bibfield  {author} {\bibinfo {author} {\bibfnamefont {K.~G.}\ \bibnamefont {Wilson}},\ }\bibfield  {title} {\bibinfo {title} {{The renormalization group: Critical phenomena and the Kondo problem}},\ }\href {https://doi.org/10.1103/RevModPhys.47.773} {\bibfield  {journal} {\bibinfo  {journal} {Rev. Mod. Phys.}\ }\textbf {\bibinfo {volume} {47}},\ \bibinfo {pages} {773} (\bibinfo {year} {1975})}\BibitemShut {NoStop}%
\bibitem [{\citenamefont {Stanley}(1999)}]{Stanley1999}%
  \BibitemOpen
  \bibfield  {author} {\bibinfo {author} {\bibfnamefont {H.~E.}\ \bibnamefont {Stanley}},\ }\bibfield  {title} {\bibinfo {title} {{Scaling, universality, and renormalization: Three pillars of modern critical phenomena}},\ }\href {https://doi.org/10.1103/RevModPhys.71.S358} {\bibfield  {journal} {\bibinfo  {journal} {Rev. Mod. Phys.}\ }\textbf {\bibinfo {volume} {71}},\ \bibinfo {pages} {S358} (\bibinfo {year} {1999})}\BibitemShut {NoStop}%
\bibitem [{\citenamefont {Braaten}\ and\ \citenamefont {Hammer}(2006)}]{BraatenHammer2006}%
  \BibitemOpen
  \bibfield  {author} {\bibinfo {author} {\bibfnamefont {E.}~\bibnamefont {Braaten}}\ and\ \bibinfo {author} {\bibfnamefont {H.-W.}\ \bibnamefont {Hammer}},\ }\bibfield  {title} {\bibinfo {title} {{Universality in few-body systems with large scattering length}},\ }\href {https://doi.org/10.1016/j.physrep.2006.03.001} {\bibfield  {journal} {\bibinfo  {journal} {Phys. Rep.}\ }\textbf {\bibinfo {volume} {428}},\ \bibinfo {pages} {259} (\bibinfo {year} {2006})}\BibitemShut {NoStop}%
\bibitem [{\citenamefont {Naidon}\ and\ \citenamefont {Endo}(2017)}]{NaidonEndo2017}%
  \BibitemOpen
  \bibfield  {author} {\bibinfo {author} {\bibfnamefont {P.}~\bibnamefont {Naidon}}\ and\ \bibinfo {author} {\bibfnamefont {S.}~\bibnamefont {Endo}},\ }\bibfield  {title} {\bibinfo {title} {{Efimov physics: a review}},\ }\href {https://doi.org/10.1088/1361-6633/aa50e8} {\bibfield  {journal} {\bibinfo  {journal} {Rep. Prog. Phys.}\ }\textbf {\bibinfo {volume} {80}},\ \bibinfo {pages} {056001} (\bibinfo {year} {2017})}\BibitemShut {NoStop}%
\bibitem [{\citenamefont {Efimov}(1970)}]{Efimov1970}%
  \BibitemOpen
  \bibfield  {author} {\bibinfo {author} {\bibfnamefont {V.}~\bibnamefont {Efimov}},\ }\bibfield  {title} {\bibinfo {title} {{Energy levels arising from resonant two-body forces in a three-body system}},\ }\href {https://doi.org/10.1016/0370-2693(70)90349-7} {\bibfield  {journal} {\bibinfo  {journal} {Phys. Lett. B}\ }\textbf {\bibinfo {volume} {33}},\ \bibinfo {pages} {563} (\bibinfo {year} {1970})}\BibitemShut {NoStop}%
\bibitem [{\citenamefont {Bedaque}\ \emph {et~al.}(1999)\citenamefont {Bedaque}, \citenamefont {Hammer},\ and\ \citenamefont {van Kolck}}]{Bedaque1999}%
  \BibitemOpen
  \bibfield  {author} {\bibinfo {author} {\bibfnamefont {P.~F.}\ \bibnamefont {Bedaque}}, \bibinfo {author} {\bibfnamefont {H.-W.}\ \bibnamefont {Hammer}},\ and\ \bibinfo {author} {\bibfnamefont {U.}~\bibnamefont {van Kolck}},\ }\bibfield  {title} {\bibinfo {title} {{Renormalization of the Three-Body System with Short-Range Interactions}},\ }\href {https://doi.org/10.1103/PhysRevLett.82.463} {\bibfield  {journal} {\bibinfo  {journal} {Phys. Rev. Lett.}\ }\textbf {\bibinfo {volume} {82}},\ \bibinfo {pages} {463} (\bibinfo {year} {1999})}\BibitemShut {NoStop}%
\bibitem [{\citenamefont {Chin}\ \emph {et~al.}(2010)\citenamefont {Chin}, \citenamefont {Grimm}, \citenamefont {Julienne},\ and\ \citenamefont {Tiesinga}}]{Chin2010}%
  \BibitemOpen
  \bibfield  {author} {\bibinfo {author} {\bibfnamefont {C.}~\bibnamefont {Chin}}, \bibinfo {author} {\bibfnamefont {R.}~\bibnamefont {Grimm}}, \bibinfo {author} {\bibfnamefont {P.}~\bibnamefont {Julienne}},\ and\ \bibinfo {author} {\bibfnamefont {E.}~\bibnamefont {Tiesinga}},\ }\bibfield  {title} {\bibinfo {title} {{Feshbach resonances in ultracold gases}},\ }\href {https://doi.org/10.1103/RevModPhys.82.1225} {\bibfield  {journal} {\bibinfo  {journal} {Rev. Mod. Phys.}\ }\textbf {\bibinfo {volume} {82}},\ \bibinfo {pages} {1225} (\bibinfo {year} {2010})}\BibitemShut {NoStop}%
\bibitem [{\citenamefont {Kraemer}\ \emph {et~al.}(2006)\citenamefont {Kraemer}, \citenamefont {Mark}, \citenamefont {Waldburger}, \citenamefont {Danzl}, \citenamefont {Chin}, \citenamefont {Engeser}, \citenamefont {Lange}, \citenamefont {Pilch}, \citenamefont {Jaakkola}, \citenamefont {N{\"{a}}gerl},\ and\ \citenamefont {Grimm}}]{Kraemer2006}%
  \BibitemOpen
  \bibfield  {author} {\bibinfo {author} {\bibfnamefont {T.}~\bibnamefont {Kraemer}}, \bibinfo {author} {\bibfnamefont {M.}~\bibnamefont {Mark}}, \bibinfo {author} {\bibfnamefont {P.}~\bibnamefont {Waldburger}}, \bibinfo {author} {\bibfnamefont {J.~G.}\ \bibnamefont {Danzl}}, \bibinfo {author} {\bibfnamefont {C.}~\bibnamefont {Chin}}, \bibinfo {author} {\bibfnamefont {B.}~\bibnamefont {Engeser}}, \bibinfo {author} {\bibfnamefont {A.~D.}\ \bibnamefont {Lange}}, \bibinfo {author} {\bibfnamefont {K.}~\bibnamefont {Pilch}}, \bibinfo {author} {\bibfnamefont {A.}~\bibnamefont {Jaakkola}}, \bibinfo {author} {\bibfnamefont {H.-C.}\ \bibnamefont {N{\"{a}}gerl}},\ and\ \bibinfo {author} {\bibfnamefont {R.}~\bibnamefont {Grimm}},\ }\bibfield  {title} {\bibinfo {title} {{Evidence for Efimov quantum states in an ultracold gas of caesium atoms}},\ }\href {https://doi.org/10.1038/nature04626} {\bibfield  {journal} {\bibinfo  {journal} {Nature}\ }\textbf {\bibinfo {volume} {440}},\ \bibinfo {pages} {315} (\bibinfo {year}
  {2006})}\BibitemShut {NoStop}%
\bibitem [{\citenamefont {Zaccanti}\ \emph {et~al.}(2009)\citenamefont {Zaccanti}, \citenamefont {Deissler}, \citenamefont {D'Errico}, \citenamefont {Fattori}, \citenamefont {Jona-Lasinio}, \citenamefont {M{\"{u}}ller}, \citenamefont {Roati}, \citenamefont {Inguscio},\ and\ \citenamefont {Modugno}}]{Zaccanti2009}%
  \BibitemOpen
  \bibfield  {author} {\bibinfo {author} {\bibfnamefont {M.}~\bibnamefont {Zaccanti}}, \bibinfo {author} {\bibfnamefont {B.}~\bibnamefont {Deissler}}, \bibinfo {author} {\bibfnamefont {C.}~\bibnamefont {D'Errico}}, \bibinfo {author} {\bibfnamefont {M.}~\bibnamefont {Fattori}}, \bibinfo {author} {\bibfnamefont {M.}~\bibnamefont {Jona-Lasinio}}, \bibinfo {author} {\bibfnamefont {S.}~\bibnamefont {M{\"{u}}ller}}, \bibinfo {author} {\bibfnamefont {G.}~\bibnamefont {Roati}}, \bibinfo {author} {\bibfnamefont {M.}~\bibnamefont {Inguscio}},\ and\ \bibinfo {author} {\bibfnamefont {G.}~\bibnamefont {Modugno}},\ }\bibfield  {title} {\bibinfo {title} {{Observation of an Efimov spectrum in an atomic system}},\ }\href {https://doi.org/10.1038/nphys1334} {\bibfield  {journal} {\bibinfo  {journal} {Nat. Phys.}\ }\textbf {\bibinfo {volume} {5}},\ \bibinfo {pages} {586} (\bibinfo {year} {2009})}\BibitemShut {NoStop}%
\bibitem [{\citenamefont {Pollack}\ \emph {et~al.}(2009)\citenamefont {Pollack}, \citenamefont {Dries},\ and\ \citenamefont {Hulet}}]{Pollack2009}%
  \BibitemOpen
  \bibfield  {author} {\bibinfo {author} {\bibfnamefont {S.~E.}\ \bibnamefont {Pollack}}, \bibinfo {author} {\bibfnamefont {D.}~\bibnamefont {Dries}},\ and\ \bibinfo {author} {\bibfnamefont {R.~G.}\ \bibnamefont {Hulet}},\ }\bibfield  {title} {\bibinfo {title} {{Universality in Three- and Four-Body Bound States of Ultracold Atoms}},\ }\href {https://doi.org/10.1126/science.1182840} {\bibfield  {journal} {\bibinfo  {journal} {Science}\ }\textbf {\bibinfo {volume} {326}},\ \bibinfo {pages} {1683} (\bibinfo {year} {2009})}\BibitemShut {NoStop}%
\bibitem [{\citenamefont {Huang}\ \emph {et~al.}(2014)\citenamefont {Huang}, \citenamefont {Sidorenkov}, \citenamefont {Grimm},\ and\ \citenamefont {Hutson}}]{Huang2014}%
  \BibitemOpen
  \bibfield  {author} {\bibinfo {author} {\bibfnamefont {B.}~\bibnamefont {Huang}}, \bibinfo {author} {\bibfnamefont {L.~A.}\ \bibnamefont {Sidorenkov}}, \bibinfo {author} {\bibfnamefont {R.}~\bibnamefont {Grimm}},\ and\ \bibinfo {author} {\bibfnamefont {J.~M.}\ \bibnamefont {Hutson}},\ }\bibfield  {title} {\bibinfo {title} {{Observation of the Second Triatomic Resonance in Efimov's Scenario}},\ }\href {https://doi.org/10.1103/PhysRevLett.112.190401} {\bibfield  {journal} {\bibinfo  {journal} {Phys. Rev. Lett.}\ }\textbf {\bibinfo {volume} {112}},\ \bibinfo {pages} {190401} (\bibinfo {year} {2014})}\BibitemShut {NoStop}%
\bibitem [{\citenamefont {Pires}\ \emph {et~al.}(2014)\citenamefont {Pires}, \citenamefont {Ulmanis}, \citenamefont {H{\"{a}}fner}, \citenamefont {Repp}, \citenamefont {Arias}, \citenamefont {Kuhnle},\ and\ \citenamefont {Weidem{\"{u}}ller}}]{Pires2014}%
  \BibitemOpen
  \bibfield  {author} {\bibinfo {author} {\bibfnamefont {R.}~\bibnamefont {Pires}}, \bibinfo {author} {\bibfnamefont {J.}~\bibnamefont {Ulmanis}}, \bibinfo {author} {\bibfnamefont {S.}~\bibnamefont {H{\"{a}}fner}}, \bibinfo {author} {\bibfnamefont {M.}~\bibnamefont {Repp}}, \bibinfo {author} {\bibfnamefont {A.}~\bibnamefont {Arias}}, \bibinfo {author} {\bibfnamefont {E.}~\bibnamefont {Kuhnle}},\ and\ \bibinfo {author} {\bibfnamefont {M.}~\bibnamefont {Weidem{\"{u}}ller}},\ }\bibfield  {title} {\bibinfo {title} {{Observation of Efimov Resonances in a Mixture with Extreme Mass Imbalance}},\ }\href {https://doi.org/10.1103/PhysRevLett.112.250404} {\bibfield  {journal} {\bibinfo  {journal} {Phys. Rev. Lett.}\ }\textbf {\bibinfo {volume} {112}},\ \bibinfo {pages} {250404} (\bibinfo {year} {2014})}\BibitemShut {NoStop}%
\bibitem [{\citenamefont {Tung}\ \emph {et~al.}(2014)\citenamefont {Tung}, \citenamefont {Jim{\'{e}}nez-Garc{\'{i}}a}, \citenamefont {Johansen}, \citenamefont {Parker},\ and\ \citenamefont {Chin}}]{Tung2014}%
  \BibitemOpen
  \bibfield  {author} {\bibinfo {author} {\bibfnamefont {S.-K.}\ \bibnamefont {Tung}}, \bibinfo {author} {\bibfnamefont {K.}~\bibnamefont {Jim{\'{e}}nez-Garc{\'{i}}a}}, \bibinfo {author} {\bibfnamefont {J.}~\bibnamefont {Johansen}}, \bibinfo {author} {\bibfnamefont {C.~V.}\ \bibnamefont {Parker}},\ and\ \bibinfo {author} {\bibfnamefont {C.}~\bibnamefont {Chin}},\ }\bibfield  {title} {\bibinfo {title} {{Geometric Scaling of Efimov States in a $^{6}\mathrm{Li}$--$^{133}\mathrm{Cs}$ Mixture}},\ }\href {https://doi.org/10.1103/PhysRevLett.113.240402} {\bibfield  {journal} {\bibinfo  {journal} {Phys. Rev. Lett.}\ }\textbf {\bibinfo {volume} {113}},\ \bibinfo {pages} {240402} (\bibinfo {year} {2014})}\BibitemShut {NoStop}%
\bibitem [{\citenamefont {Porras}\ and\ \citenamefont {Cirac}(2004)}]{Porras2004}%
  \BibitemOpen
  \bibfield  {author} {\bibinfo {author} {\bibfnamefont {D.}~\bibnamefont {Porras}}\ and\ \bibinfo {author} {\bibfnamefont {J.~I.}\ \bibnamefont {Cirac}},\ }\bibfield  {title} {\bibinfo {title} {{Effective Quantum Spin Systems with Trapped Ions}},\ }\href {https://doi.org/10.1103/physrevlett.92.207901} {\bibfield  {journal} {\bibinfo  {journal} {Phys. Rev. Lett.}\ }\textbf {\bibinfo {volume} {92}},\ \bibinfo {pages} {207901} (\bibinfo {year} {2004})}\BibitemShut {NoStop}%
\bibitem [{\citenamefont {Kim}\ \emph {et~al.}(2009)\citenamefont {Kim}, \citenamefont {Chang}, \citenamefont {Islam}, \citenamefont {Korenblit}, \citenamefont {Duan},\ and\ \citenamefont {Monroe}}]{Kim2009}%
  \BibitemOpen
  \bibfield  {author} {\bibinfo {author} {\bibfnamefont {K.}~\bibnamefont {Kim}}, \bibinfo {author} {\bibfnamefont {M.-S.}\ \bibnamefont {Chang}}, \bibinfo {author} {\bibfnamefont {R.}~\bibnamefont {Islam}}, \bibinfo {author} {\bibfnamefont {S.}~\bibnamefont {Korenblit}}, \bibinfo {author} {\bibfnamefont {L.-M.}\ \bibnamefont {Duan}},\ and\ \bibinfo {author} {\bibfnamefont {C.}~\bibnamefont {Monroe}},\ }\bibfield  {title} {\bibinfo {title} {{Entanglement and Tunable Spin-Spin Couplings between Trapped Ions Using Multiple Transverse Modes}},\ }\href {https://doi.org/10.1103/physrevlett.103.120502} {\bibfield  {journal} {\bibinfo  {journal} {Phys. Rev. Lett.}\ }\textbf {\bibinfo {volume} {103}},\ \bibinfo {pages} {120502} (\bibinfo {year} {2009})}\BibitemShut {NoStop}%
\bibitem [{\citenamefont {Britton}\ \emph {et~al.}(2012)\citenamefont {Britton}, \citenamefont {Sawyer}, \citenamefont {Keith}, \citenamefont {Wang}, \citenamefont {Freericks}, \citenamefont {Uys}, \citenamefont {Biercuk},\ and\ \citenamefont {Bollinger}}]{Britton2012}%
  \BibitemOpen
  \bibfield  {author} {\bibinfo {author} {\bibfnamefont {J.~W.}\ \bibnamefont {Britton}}, \bibinfo {author} {\bibfnamefont {B.~C.}\ \bibnamefont {Sawyer}}, \bibinfo {author} {\bibfnamefont {A.~C.}\ \bibnamefont {Keith}}, \bibinfo {author} {\bibfnamefont {C.-C.~J.}\ \bibnamefont {Wang}}, \bibinfo {author} {\bibfnamefont {J.~K.}\ \bibnamefont {Freericks}}, \bibinfo {author} {\bibfnamefont {H.}~\bibnamefont {Uys}}, \bibinfo {author} {\bibfnamefont {M.~J.}\ \bibnamefont {Biercuk}},\ and\ \bibinfo {author} {\bibfnamefont {J.~J.}\ \bibnamefont {Bollinger}},\ }\bibfield  {title} {\bibinfo {title} {{Engineered two-dimensional Ising interactions in a trapped-ion quantum simulator with hundreds of spins}},\ }\href {https://doi.org/10.1038/nature10981} {\bibfield  {journal} {\bibinfo  {journal} {Nature}\ }\textbf {\bibinfo {volume} {484}},\ \bibinfo {pages} {489} (\bibinfo {year} {2012})}\BibitemShut {NoStop}%
\bibitem [{\citenamefont {Monroe}\ \emph {et~al.}(2021)\citenamefont {Monroe}, \citenamefont {Campbell}, \citenamefont {Duan}, \citenamefont {Gong}, \citenamefont {Gorshkov}, \citenamefont {Hess}, \citenamefont {Islam}, \citenamefont {Kim}, \citenamefont {Linke}, \citenamefont {Pagano}, \citenamefont {Richerme}, \citenamefont {Senko},\ and\ \citenamefont {Yao}}]{Monroe2021}%
  \BibitemOpen
  \bibfield  {author} {\bibinfo {author} {\bibfnamefont {C.}~\bibnamefont {Monroe}}, \bibinfo {author} {\bibfnamefont {W.~C.}\ \bibnamefont {Campbell}}, \bibinfo {author} {\bibfnamefont {L.-M.}\ \bibnamefont {Duan}}, \bibinfo {author} {\bibfnamefont {Z.-X.}\ \bibnamefont {Gong}}, \bibinfo {author} {\bibfnamefont {A.~V.}\ \bibnamefont {Gorshkov}}, \bibinfo {author} {\bibfnamefont {P.~W.}\ \bibnamefont {Hess}}, \bibinfo {author} {\bibfnamefont {R.}~\bibnamefont {Islam}}, \bibinfo {author} {\bibfnamefont {K.}~\bibnamefont {Kim}}, \bibinfo {author} {\bibfnamefont {N.~M.}\ \bibnamefont {Linke}}, \bibinfo {author} {\bibfnamefont {G.}~\bibnamefont {Pagano}}, \bibinfo {author} {\bibfnamefont {P.}~\bibnamefont {Richerme}}, \bibinfo {author} {\bibfnamefont {C.}~\bibnamefont {Senko}},\ and\ \bibinfo {author} {\bibfnamefont {N.~Y.}\ \bibnamefont {Yao}},\ }\bibfield  {title} {\bibinfo {title} {{Programmable quantum simulations of spin systems with trapped ions}},\ }\href {https://doi.org/10.1103/revmodphys.93.025001}
  {\bibfield  {journal} {\bibinfo  {journal} {Rev. Mod. Phys.}\ }\textbf {\bibinfo {volume} {93}},\ \bibinfo {pages} {025001} (\bibinfo {year} {2021})}\BibitemShut {NoStop}%
\bibitem [{\citenamefont {Jurcevic}\ \emph {et~al.}(2014)\citenamefont {Jurcevic}, \citenamefont {Lanyon}, \citenamefont {Hauke}, \citenamefont {Hempel}, \citenamefont {Zoller}, \citenamefont {Blatt},\ and\ \citenamefont {Roos}}]{Jurcevic2014}%
  \BibitemOpen
  \bibfield  {author} {\bibinfo {author} {\bibfnamefont {P.}~\bibnamefont {Jurcevic}}, \bibinfo {author} {\bibfnamefont {B.~P.}\ \bibnamefont {Lanyon}}, \bibinfo {author} {\bibfnamefont {P.}~\bibnamefont {Hauke}}, \bibinfo {author} {\bibfnamefont {C.}~\bibnamefont {Hempel}}, \bibinfo {author} {\bibfnamefont {P.}~\bibnamefont {Zoller}}, \bibinfo {author} {\bibfnamefont {R.}~\bibnamefont {Blatt}},\ and\ \bibinfo {author} {\bibfnamefont {C.~F.}\ \bibnamefont {Roos}},\ }\bibfield  {title} {\bibinfo {title} {{Quasiparticle engineering and entanglement propagation in a quantum many-body system}},\ }\href {https://doi.org/10.1038/nature13461} {\bibfield  {journal} {\bibinfo  {journal} {Nature}\ }\textbf {\bibinfo {volume} {511}},\ \bibinfo {pages} {202} (\bibinfo {year} {2014})}\BibitemShut {NoStop}%
\bibitem [{\citenamefont {Richerme}\ \emph {et~al.}(2014)\citenamefont {Richerme}, \citenamefont {Gong}, \citenamefont {Lee}, \citenamefont {Senko}, \citenamefont {Smith}, \citenamefont {Foss-Feig}, \citenamefont {Michalakis}, \citenamefont {Gorshkov},\ and\ \citenamefont {Monroe}}]{Richerme2014}%
  \BibitemOpen
  \bibfield  {author} {\bibinfo {author} {\bibfnamefont {P.}~\bibnamefont {Richerme}}, \bibinfo {author} {\bibfnamefont {Z.-X.}\ \bibnamefont {Gong}}, \bibinfo {author} {\bibfnamefont {A.}~\bibnamefont {Lee}}, \bibinfo {author} {\bibfnamefont {C.}~\bibnamefont {Senko}}, \bibinfo {author} {\bibfnamefont {J.}~\bibnamefont {Smith}}, \bibinfo {author} {\bibfnamefont {M.}~\bibnamefont {Foss-Feig}}, \bibinfo {author} {\bibfnamefont {S.}~\bibnamefont {Michalakis}}, \bibinfo {author} {\bibfnamefont {A.~V.}\ \bibnamefont {Gorshkov}},\ and\ \bibinfo {author} {\bibfnamefont {C.}~\bibnamefont {Monroe}},\ }\bibfield  {title} {\bibinfo {title} {{Non-local propagation of correlations in quantum systems with long-range interactions}},\ }\href {https://doi.org/10.1038/nature13450} {\bibfield  {journal} {\bibinfo  {journal} {Nature}\ }\textbf {\bibinfo {volume} {511}},\ \bibinfo {pages} {198} (\bibinfo {year} {2014})}\BibitemShut {NoStop}%
\bibitem [{\citenamefont {Senko}\ \emph {et~al.}(2014)\citenamefont {Senko}, \citenamefont {Smith}, \citenamefont {Richerme}, \citenamefont {Lee}, \citenamefont {Campbell},\ and\ \citenamefont {Monroe}}]{Senko2014}%
  \BibitemOpen
  \bibfield  {author} {\bibinfo {author} {\bibfnamefont {C.}~\bibnamefont {Senko}}, \bibinfo {author} {\bibfnamefont {J.}~\bibnamefont {Smith}}, \bibinfo {author} {\bibfnamefont {P.}~\bibnamefont {Richerme}}, \bibinfo {author} {\bibfnamefont {A.}~\bibnamefont {Lee}}, \bibinfo {author} {\bibfnamefont {W.~C.}\ \bibnamefont {Campbell}},\ and\ \bibinfo {author} {\bibfnamefont {C.}~\bibnamefont {Monroe}},\ }\bibfield  {title} {\bibinfo {title} {{Coherent imaging spectroscopy of a quantum many-body spin system}},\ }\href {https://doi.org/10.1126/science.1251422} {\bibfield  {journal} {\bibinfo  {journal} {Science}\ }\textbf {\bibinfo {volume} {345}},\ \bibinfo {pages} {430} (\bibinfo {year} {2014})}\BibitemShut {NoStop}%
\bibitem [{\citenamefont {Guo}\ \emph {et~al.}(2024)\citenamefont {Guo}, \citenamefont {Wu}, \citenamefont {Ye}, \citenamefont {Zhang}, \citenamefont {Lian}, \citenamefont {Yao}, \citenamefont {Wang}, \citenamefont {Yan}, \citenamefont {Yi}, \citenamefont {Xu}, \citenamefont {Li}, \citenamefont {Hou}, \citenamefont {Xu}, \citenamefont {Guo}, \citenamefont {Zhang}, \citenamefont {Qi}, \citenamefont {Zhou}, \citenamefont {He},\ and\ \citenamefont {Duan}}]{Guo2024}%
  \BibitemOpen
  \bibfield  {author} {\bibinfo {author} {\bibfnamefont {S.-A.}\ \bibnamefont {Guo}}, \bibinfo {author} {\bibfnamefont {Y.-K.}\ \bibnamefont {Wu}}, \bibinfo {author} {\bibfnamefont {J.}~\bibnamefont {Ye}}, \bibinfo {author} {\bibfnamefont {L.}~\bibnamefont {Zhang}}, \bibinfo {author} {\bibfnamefont {W.-Q.}\ \bibnamefont {Lian}}, \bibinfo {author} {\bibfnamefont {R.}~\bibnamefont {Yao}}, \bibinfo {author} {\bibfnamefont {Y.}~\bibnamefont {Wang}}, \bibinfo {author} {\bibfnamefont {R.-Y.}\ \bibnamefont {Yan}}, \bibinfo {author} {\bibfnamefont {Y.-J.}\ \bibnamefont {Yi}}, \bibinfo {author} {\bibfnamefont {Y.-L.}\ \bibnamefont {Xu}}, \bibinfo {author} {\bibfnamefont {B.-W.}\ \bibnamefont {Li}}, \bibinfo {author} {\bibfnamefont {Y.-H.}\ \bibnamefont {Hou}}, \bibinfo {author} {\bibfnamefont {Y.-Z.}\ \bibnamefont {Xu}}, \bibinfo {author} {\bibfnamefont {W.-X.}\ \bibnamefont {Guo}}, \bibinfo {author} {\bibfnamefont {C.}~\bibnamefont {Zhang}}, \bibinfo {author} {\bibfnamefont {B.-X.}\ \bibnamefont {Qi}}, \bibinfo
  {author} {\bibfnamefont {Z.-C.}\ \bibnamefont {Zhou}}, \bibinfo {author} {\bibfnamefont {L.}~\bibnamefont {He}},\ and\ \bibinfo {author} {\bibfnamefont {L.-M.}\ \bibnamefont {Duan}},\ }\bibfield  {title} {\bibinfo {title} {A site-resolved two-dimensional quantum simulator with hundreds of trapped ions},\ }\href {https://doi.org/10.1038/s41586-024-07459-0} {\bibfield  {journal} {\bibinfo  {journal} {Nature}\ }\textbf {\bibinfo {volume} {630}},\ \bibinfo {pages} {613} (\bibinfo {year} {2024})}\BibitemShut {NoStop}%
\bibitem [{\citenamefont {Kim}\ \emph {et~al.}(2010)\citenamefont {Kim}, \citenamefont {Chang}, \citenamefont {Korenblit}, \citenamefont {Islam}, \citenamefont {Edwards}, \citenamefont {Freericks}, \citenamefont {Lin}, \citenamefont {Duan},\ and\ \citenamefont {Monroe}}]{Kim2010}%
  \BibitemOpen
  \bibfield  {author} {\bibinfo {author} {\bibfnamefont {K.}~\bibnamefont {Kim}}, \bibinfo {author} {\bibfnamefont {M.-S.}\ \bibnamefont {Chang}}, \bibinfo {author} {\bibfnamefont {S.}~\bibnamefont {Korenblit}}, \bibinfo {author} {\bibfnamefont {R.}~\bibnamefont {Islam}}, \bibinfo {author} {\bibfnamefont {E.~E.}\ \bibnamefont {Edwards}}, \bibinfo {author} {\bibfnamefont {J.~K.}\ \bibnamefont {Freericks}}, \bibinfo {author} {\bibfnamefont {G.-D.}\ \bibnamefont {Lin}}, \bibinfo {author} {\bibfnamefont {L.-M.}\ \bibnamefont {Duan}},\ and\ \bibinfo {author} {\bibfnamefont {C.}~\bibnamefont {Monroe}},\ }\bibfield  {title} {\bibinfo {title} {{Quantum simulation of frustrated Ising spins with trapped ions}},\ }\href {https://doi.org/10.1038/nature09071} {\bibfield  {journal} {\bibinfo  {journal} {Nature}\ }\textbf {\bibinfo {volume} {465}},\ \bibinfo {pages} {590} (\bibinfo {year} {2010})}\BibitemShut {NoStop}%
\bibitem [{\citenamefont {Islam}\ \emph {et~al.}(2013)\citenamefont {Islam}, \citenamefont {Senko}, \citenamefont {Campbell}, \citenamefont {Korenblit}, \citenamefont {Smith}, \citenamefont {Lee}, \citenamefont {Edwards}, \citenamefont {Wang}, \citenamefont {Freericks},\ and\ \citenamefont {Monroe}}]{Islam2013}%
  \BibitemOpen
  \bibfield  {author} {\bibinfo {author} {\bibfnamefont {R.}~\bibnamefont {Islam}}, \bibinfo {author} {\bibfnamefont {C.}~\bibnamefont {Senko}}, \bibinfo {author} {\bibfnamefont {W.~C.}\ \bibnamefont {Campbell}}, \bibinfo {author} {\bibfnamefont {S.}~\bibnamefont {Korenblit}}, \bibinfo {author} {\bibfnamefont {J.}~\bibnamefont {Smith}}, \bibinfo {author} {\bibfnamefont {A.}~\bibnamefont {Lee}}, \bibinfo {author} {\bibfnamefont {E.~E.}\ \bibnamefont {Edwards}}, \bibinfo {author} {\bibfnamefont {C.-C.~J.}\ \bibnamefont {Wang}}, \bibinfo {author} {\bibfnamefont {J.~K.}\ \bibnamefont {Freericks}},\ and\ \bibinfo {author} {\bibfnamefont {C.}~\bibnamefont {Monroe}},\ }\bibfield  {title} {\bibinfo {title} {{Emergence and Frustration of Magnetism with Variable-Range Interactions in a Quantum Simulator}},\ }\href {https://doi.org/10.1126/science.1232296} {\bibfield  {journal} {\bibinfo  {journal} {Science}\ }\textbf {\bibinfo {volume} {340}},\ \bibinfo {pages} {583} (\bibinfo {year} {2013})}\BibitemShut {NoStop}%
\bibitem [{\citenamefont {Feng}\ \emph {et~al.}(2023)\citenamefont {Feng}, \citenamefont {Katz}, \citenamefont {Haack}, \citenamefont {Maghrebi}, \citenamefont {Gorshkov}, \citenamefont {Gong}, \citenamefont {Cetina},\ and\ \citenamefont {Monroe}}]{Feng2023}%
  \BibitemOpen
  \bibfield  {author} {\bibinfo {author} {\bibfnamefont {L.}~\bibnamefont {Feng}}, \bibinfo {author} {\bibfnamefont {O.}~\bibnamefont {Katz}}, \bibinfo {author} {\bibfnamefont {C.}~\bibnamefont {Haack}}, \bibinfo {author} {\bibfnamefont {M.}~\bibnamefont {Maghrebi}}, \bibinfo {author} {\bibfnamefont {A.~V.}\ \bibnamefont {Gorshkov}}, \bibinfo {author} {\bibfnamefont {Z.}~\bibnamefont {Gong}}, \bibinfo {author} {\bibfnamefont {M.}~\bibnamefont {Cetina}},\ and\ \bibinfo {author} {\bibfnamefont {C.}~\bibnamefont {Monroe}},\ }\bibfield  {title} {\bibinfo {title} {{Continuous symmetry breaking in a trapped-ion spin chain}},\ }\href {https://doi.org/10.1038/s41586-023-06656-7} {\bibfield  {journal} {\bibinfo  {journal} {Nature}\ }\textbf {\bibinfo {volume} {623}},\ \bibinfo {pages} {713} (\bibinfo {year} {2023})}\BibitemShut {NoStop}%
\bibitem [{\citenamefont {Smith}\ \emph {et~al.}(2016)\citenamefont {Smith}, \citenamefont {Lee}, \citenamefont {Richerme}, \citenamefont {Neyenhuis}, \citenamefont {Hess}, \citenamefont {Hauke}, \citenamefont {Heyl}, \citenamefont {Huse},\ and\ \citenamefont {Monroe}}]{Smith2016}%
  \BibitemOpen
  \bibfield  {author} {\bibinfo {author} {\bibfnamefont {J.}~\bibnamefont {Smith}}, \bibinfo {author} {\bibfnamefont {A.}~\bibnamefont {Lee}}, \bibinfo {author} {\bibfnamefont {P.}~\bibnamefont {Richerme}}, \bibinfo {author} {\bibfnamefont {B.}~\bibnamefont {Neyenhuis}}, \bibinfo {author} {\bibfnamefont {P.~W.}\ \bibnamefont {Hess}}, \bibinfo {author} {\bibfnamefont {P.}~\bibnamefont {Hauke}}, \bibinfo {author} {\bibfnamefont {M.}~\bibnamefont {Heyl}}, \bibinfo {author} {\bibfnamefont {D.~A.}\ \bibnamefont {Huse}},\ and\ \bibinfo {author} {\bibfnamefont {C.}~\bibnamefont {Monroe}},\ }\bibfield  {title} {\bibinfo {title} {{Many-body localization in a quantum simulator with programmable random disorder}},\ }\href {https://doi.org/10.1038/nphys3783} {\bibfield  {journal} {\bibinfo  {journal} {Nat. Phys.}\ }\textbf {\bibinfo {volume} {12}},\ \bibinfo {pages} {907} (\bibinfo {year} {2016})}\BibitemShut {NoStop}%
\bibitem [{\citenamefont {Zhang}\ \emph {et~al.}(2017)\citenamefont {Zhang}, \citenamefont {Pagano}, \citenamefont {Hess}, \citenamefont {Kyprianidis}, \citenamefont {Becker}, \citenamefont {Kaplan}, \citenamefont {Gorshkov}, \citenamefont {Gong},\ and\ \citenamefont {Monroe}}]{Zhang2017Dynamics}%
  \BibitemOpen
  \bibfield  {author} {\bibinfo {author} {\bibfnamefont {J.}~\bibnamefont {Zhang}}, \bibinfo {author} {\bibfnamefont {G.}~\bibnamefont {Pagano}}, \bibinfo {author} {\bibfnamefont {P.~W.}\ \bibnamefont {Hess}}, \bibinfo {author} {\bibfnamefont {A.}~\bibnamefont {Kyprianidis}}, \bibinfo {author} {\bibfnamefont {P.}~\bibnamefont {Becker}}, \bibinfo {author} {\bibfnamefont {H.}~\bibnamefont {Kaplan}}, \bibinfo {author} {\bibfnamefont {A.~V.}\ \bibnamefont {Gorshkov}}, \bibinfo {author} {\bibfnamefont {Z.-X.}\ \bibnamefont {Gong}},\ and\ \bibinfo {author} {\bibfnamefont {C.}~\bibnamefont {Monroe}},\ }\bibfield  {title} {\bibinfo {title} {{Observation of a many-body dynamical phase transition with a 53-qubit quantum simulator}},\ }\href {https://doi.org/10.1038/nature24654} {\bibfield  {journal} {\bibinfo  {journal} {Nature}\ }\textbf {\bibinfo {volume} {551}},\ \bibinfo {pages} {601} (\bibinfo {year} {2017})}\BibitemShut {NoStop}%
\bibitem [{\citenamefont {Morong}\ \emph {et~al.}(2021)\citenamefont {Morong}, \citenamefont {Liu}, \citenamefont {Becker}, \citenamefont {Collins}, \citenamefont {Feng}, \citenamefont {Kyprianidis}, \citenamefont {Pagano}, \citenamefont {You}, \citenamefont {Gorshkov},\ and\ \citenamefont {Monroe}}]{Morong2021}%
  \BibitemOpen
  \bibfield  {author} {\bibinfo {author} {\bibfnamefont {W.}~\bibnamefont {Morong}}, \bibinfo {author} {\bibfnamefont {F.}~\bibnamefont {Liu}}, \bibinfo {author} {\bibfnamefont {P.}~\bibnamefont {Becker}}, \bibinfo {author} {\bibfnamefont {K.~S.}\ \bibnamefont {Collins}}, \bibinfo {author} {\bibfnamefont {L.}~\bibnamefont {Feng}}, \bibinfo {author} {\bibfnamefont {A.}~\bibnamefont {Kyprianidis}}, \bibinfo {author} {\bibfnamefont {G.}~\bibnamefont {Pagano}}, \bibinfo {author} {\bibfnamefont {T.}~\bibnamefont {You}}, \bibinfo {author} {\bibfnamefont {A.~V.}\ \bibnamefont {Gorshkov}},\ and\ \bibinfo {author} {\bibfnamefont {C.}~\bibnamefont {Monroe}},\ }\bibfield  {title} {\bibinfo {title} {{Observation of Stark many-body localization without disorder}},\ }\href {https://doi.org/10.1038/s41586-021-03988-0} {\bibfield  {journal} {\bibinfo  {journal} {Nature}\ }\textbf {\bibinfo {volume} {599}},\ \bibinfo {pages} {393} (\bibinfo {year} {2021})}\BibitemShut {NoStop}%
\bibitem [{\citenamefont {Brydges}\ \emph {et~al.}(2019)\citenamefont {Brydges}, \citenamefont {Elben}, \citenamefont {Jurcevic}, \citenamefont {Vermersch}, \citenamefont {Maier}, \citenamefont {Lanyon}, \citenamefont {Zoller}, \citenamefont {Blatt},\ and\ \citenamefont {Roos}}]{Brydges2019}%
  \BibitemOpen
  \bibfield  {author} {\bibinfo {author} {\bibfnamefont {T.}~\bibnamefont {Brydges}}, \bibinfo {author} {\bibfnamefont {A.}~\bibnamefont {Elben}}, \bibinfo {author} {\bibfnamefont {P.}~\bibnamefont {Jurcevic}}, \bibinfo {author} {\bibfnamefont {B.}~\bibnamefont {Vermersch}}, \bibinfo {author} {\bibfnamefont {C.}~\bibnamefont {Maier}}, \bibinfo {author} {\bibfnamefont {B.~P.}\ \bibnamefont {Lanyon}}, \bibinfo {author} {\bibfnamefont {P.}~\bibnamefont {Zoller}}, \bibinfo {author} {\bibfnamefont {R.}~\bibnamefont {Blatt}},\ and\ \bibinfo {author} {\bibfnamefont {C.~F.}\ \bibnamefont {Roos}},\ }\bibfield  {title} {\bibinfo {title} {{Probing R{\'{e}}nyi entanglement entropy via randomized measurements}},\ }\href {https://doi.org/10.1126/science.aau4963} {\bibfield  {journal} {\bibinfo  {journal} {Science}\ }\textbf {\bibinfo {volume} {364}},\ \bibinfo {pages} {260} (\bibinfo {year} {2019})}\BibitemShut {NoStop}%
\bibitem [{\citenamefont {Joshi}\ \emph {et~al.}(2022)\citenamefont {Joshi}, \citenamefont {Kranzl}, \citenamefont {Schuckert}, \citenamefont {Lovas}, \citenamefont {Maier}, \citenamefont {Blatt}, \citenamefont {Knap},\ and\ \citenamefont {Roos}}]{Joshi2022}%
  \BibitemOpen
  \bibfield  {author} {\bibinfo {author} {\bibfnamefont {M.~K.}\ \bibnamefont {Joshi}}, \bibinfo {author} {\bibfnamefont {F.}~\bibnamefont {Kranzl}}, \bibinfo {author} {\bibfnamefont {A.}~\bibnamefont {Schuckert}}, \bibinfo {author} {\bibfnamefont {I.}~\bibnamefont {Lovas}}, \bibinfo {author} {\bibfnamefont {C.}~\bibnamefont {Maier}}, \bibinfo {author} {\bibfnamefont {R.}~\bibnamefont {Blatt}}, \bibinfo {author} {\bibfnamefont {M.}~\bibnamefont {Knap}},\ and\ \bibinfo {author} {\bibfnamefont {C.~F.}\ \bibnamefont {Roos}},\ }\bibfield  {title} {\bibinfo {title} {{Observing emergent hydrodynamics in a long-range quantum magnet}},\ }\href {https://doi.org/10.1126/science.abk2400} {\bibfield  {journal} {\bibinfo  {journal} {Science}\ }\textbf {\bibinfo {volume} {376}},\ \bibinfo {pages} {720} (\bibinfo {year} {2022})}\BibitemShut {NoStop}%
\bibitem [{\citenamefont {Fukuhara}\ \emph {et~al.}(2013)\citenamefont {Fukuhara}, \citenamefont {Schau{\ss}}, \citenamefont {Endres}, \citenamefont {Hild}, \citenamefont {Cheneau}, \citenamefont {Bloch},\ and\ \citenamefont {Gross}}]{Fukuhara2013}%
  \BibitemOpen
  \bibfield  {author} {\bibinfo {author} {\bibfnamefont {T.}~\bibnamefont {Fukuhara}}, \bibinfo {author} {\bibfnamefont {P.}~\bibnamefont {Schau{\ss}}}, \bibinfo {author} {\bibfnamefont {M.}~\bibnamefont {Endres}}, \bibinfo {author} {\bibfnamefont {S.}~\bibnamefont {Hild}}, \bibinfo {author} {\bibfnamefont {M.}~\bibnamefont {Cheneau}}, \bibinfo {author} {\bibfnamefont {I.}~\bibnamefont {Bloch}},\ and\ \bibinfo {author} {\bibfnamefont {C.}~\bibnamefont {Gross}},\ }\bibfield  {title} {\bibinfo {title} {{Microscopic observation of magnon bound states and their dynamics}},\ }\href {https://doi.org/10.1038/nature12541} {\bibfield  {journal} {\bibinfo  {journal} {Nature}\ }\textbf {\bibinfo {volume} {502}},\ \bibinfo {pages} {76} (\bibinfo {year} {2013})}\BibitemShut {NoStop}%
\bibitem [{\citenamefont {Kranzl}\ \emph {et~al.}(2023)\citenamefont {Kranzl}, \citenamefont {Birnkammer}, \citenamefont {Joshi}, \citenamefont {Bastianello}, \citenamefont {Blatt}, \citenamefont {Knap},\ and\ \citenamefont {Roos}}]{Kranzl2023}%
  \BibitemOpen
  \bibfield  {author} {\bibinfo {author} {\bibfnamefont {F.}~\bibnamefont {Kranzl}}, \bibinfo {author} {\bibfnamefont {S.}~\bibnamefont {Birnkammer}}, \bibinfo {author} {\bibfnamefont {M.~K.}\ \bibnamefont {Joshi}}, \bibinfo {author} {\bibfnamefont {A.}~\bibnamefont {Bastianello}}, \bibinfo {author} {\bibfnamefont {R.}~\bibnamefont {Blatt}}, \bibinfo {author} {\bibfnamefont {M.}~\bibnamefont {Knap}},\ and\ \bibinfo {author} {\bibfnamefont {C.~F.}\ \bibnamefont {Roos}},\ }\bibfield  {title} {\bibinfo {title} {{Observation of Magnon Bound States in the Long-Range, Anisotropic Heisenberg Model}},\ }\href {https://doi.org/10.1103/physrevx.13.031017} {\bibfield  {journal} {\bibinfo  {journal} {Phys. Rev. X}\ }\textbf {\bibinfo {volume} {13}},\ \bibinfo {pages} {031017} (\bibinfo {year} {2023})}\BibitemShut {NoStop}%
\bibitem [{\citenamefont {Sun}\ \emph {et~al.}(2026{\natexlab{a}})\citenamefont {Sun}, \citenamefont {Feng},\ and\ \citenamefont {Zhang}}]{Sun2026Efimov}%
  \BibitemOpen
  \bibfield  {author} {\bibinfo {author} {\bibfnamefont {N.}~\bibnamefont {Sun}}, \bibinfo {author} {\bibfnamefont {L.}~\bibnamefont {Feng}},\ and\ \bibinfo {author} {\bibfnamefont {P.}~\bibnamefont {Zhang}},\ }\bibfield  {title} {\bibinfo {title} {{Efimov effect in long-range quantum spin chains}},\ }\href {https://doi.org/10.1038/s42005-026-02580-0} {\bibfield  {journal} {\bibinfo  {journal} {Commun. Phys.}\ }\textbf {\bibinfo {volume} {9}},\ \bibinfo {pages} {146} (\bibinfo {year} {2026}{\natexlab{a}})}\BibitemShut {NoStop}%
\bibitem [{\citenamefont {Sun}\ \emph {et~al.}(2026{\natexlab{b}})\citenamefont {Sun}, \citenamefont {Feng},\ and\ \citenamefont {Zhang}}]{Sun2026Impurity}%
  \BibitemOpen
  \bibfield  {author} {\bibinfo {author} {\bibfnamefont {N.}~\bibnamefont {Sun}}, \bibinfo {author} {\bibfnamefont {L.}~\bibnamefont {Feng}},\ and\ \bibinfo {author} {\bibfnamefont {P.}~\bibnamefont {Zhang}},\ }\bibfield  {title} {\bibinfo {title} {{Universal bound states in long-range spin chains with an impurity}},\ }\href {https://doi.org/10.1103/c7lm-323v} {\bibfield  {journal} {\bibinfo  {journal} {Phys. Rev. B}\ }\textbf {\bibinfo {volume} {113}},\ \bibinfo {pages} {134442} (\bibinfo {year} {2026}{\natexlab{b}})}\BibitemShut {NoStop}%
\bibitem [{\citenamefont {Sun}\ \emph {et~al.}(2026{\natexlab{c}})\citenamefont {Sun}, \citenamefont {Feng},\ and\ \citenamefont {Zhang}}]{Sun2026Contact}%
  \BibitemOpen
  \bibfield  {author} {\bibinfo {author} {\bibfnamefont {N.}~\bibnamefont {Sun}}, \bibinfo {author} {\bibfnamefont {L.}~\bibnamefont {Feng}},\ and\ \bibinfo {author} {\bibfnamefont {P.}~\bibnamefont {Zhang}},\ }\bibfield  {title} {\bibinfo {title} {{Universal relations in long-range quantum spin chains}},\ }\href {https://doi.org/10.1103/4nf5-q435} {\bibfield  {journal} {\bibinfo  {journal} {Phys. Rev. A}\ }\textbf {\bibinfo {volume} {113}},\ \bibinfo {pages} {L041308} (\bibinfo {year} {2026}{\natexlab{c}})}\BibitemShut {NoStop}%
\bibitem [{\citenamefont {Wang}\ \emph {et~al.}(2022)\citenamefont {Wang}, \citenamefont {Gullans}, \citenamefont {Na}, \citenamefont {Whitsitt},\ and\ \citenamefont {Gorshkov}}]{Wang2022}%
  \BibitemOpen
  \bibfield  {author} {\bibinfo {author} {\bibfnamefont {Y.}~\bibnamefont {Wang}}, \bibinfo {author} {\bibfnamefont {M.~J.}\ \bibnamefont {Gullans}}, \bibinfo {author} {\bibfnamefont {X.}~\bibnamefont {Na}}, \bibinfo {author} {\bibfnamefont {S.}~\bibnamefont {Whitsitt}},\ and\ \bibinfo {author} {\bibfnamefont {A.~V.}\ \bibnamefont {Gorshkov}},\ }\bibfield  {title} {\bibinfo {title} {{Universal scattering with general dispersion relations}},\ }\href {https://doi.org/10.1103/PhysRevResearch.4.023014} {\bibfield  {journal} {\bibinfo  {journal} {Phys. Rev. Research}\ }\textbf {\bibinfo {volume} {4}},\ \bibinfo {pages} {023014} (\bibinfo {year} {2022})}\BibitemShut {NoStop}%
\bibitem [{\citenamefont {Deng}\ \emph {et~al.}(2016)\citenamefont {Deng}, \citenamefont {Shi}, \citenamefont {Diao}, \citenamefont {Yu}, \citenamefont {Zhai}, \citenamefont {Qi},\ and\ \citenamefont {Wu}}]{Deng2016}%
  \BibitemOpen
  \bibfield  {author} {\bibinfo {author} {\bibfnamefont {S.}~\bibnamefont {Deng}}, \bibinfo {author} {\bibfnamefont {Z.-Y.}\ \bibnamefont {Shi}}, \bibinfo {author} {\bibfnamefont {P.}~\bibnamefont {Diao}}, \bibinfo {author} {\bibfnamefont {Q.}~\bibnamefont {Yu}}, \bibinfo {author} {\bibfnamefont {H.}~\bibnamefont {Zhai}}, \bibinfo {author} {\bibfnamefont {R.}~\bibnamefont {Qi}},\ and\ \bibinfo {author} {\bibfnamefont {H.}~\bibnamefont {Wu}},\ }\bibfield  {title} {\bibinfo {title} {{Observation of the Efimovian expansion in scale-invariant Fermi gases}},\ }\href {https://doi.org/10.1126/science.aaf0666} {\bibfield  {journal} {\bibinfo  {journal} {Science}\ }\textbf {\bibinfo {volume} {353}},\ \bibinfo {pages} {371} (\bibinfo {year} {2016})}\BibitemShut {NoStop}%
\bibitem [{\citenamefont {Shi}\ \emph {et~al.}(2017)\citenamefont {Shi}, \citenamefont {Qi}, \citenamefont {Zhai},\ and\ \citenamefont {Yu}}]{Shi2017}%
  \BibitemOpen
  \bibfield  {author} {\bibinfo {author} {\bibfnamefont {Z.-Y.}\ \bibnamefont {Shi}}, \bibinfo {author} {\bibfnamefont {R.}~\bibnamefont {Qi}}, \bibinfo {author} {\bibfnamefont {H.}~\bibnamefont {Zhai}},\ and\ \bibinfo {author} {\bibfnamefont {Z.}~\bibnamefont {Yu}},\ }\bibfield  {title} {\bibinfo {title} {{Dynamic super Efimov effect}},\ }\href {https://doi.org/10.1103/physreva.96.050702} {\bibfield  {journal} {\bibinfo  {journal} {Phys. Rev. A}\ }\textbf {\bibinfo {volume} {96}},\ \bibinfo {pages} {050702} (\bibinfo {year} {2017})}\BibitemShut {NoStop}%
\bibitem [{\citenamefont {Deng}\ \emph {et~al.}(2018)\citenamefont {Deng}, \citenamefont {Diao}, \citenamefont {Li}, \citenamefont {Yu}, \citenamefont {Yu},\ and\ \citenamefont {Wu}}]{Deng2018}%
  \BibitemOpen
  \bibfield  {author} {\bibinfo {author} {\bibfnamefont {S.}~\bibnamefont {Deng}}, \bibinfo {author} {\bibfnamefont {P.}~\bibnamefont {Diao}}, \bibinfo {author} {\bibfnamefont {F.}~\bibnamefont {Li}}, \bibinfo {author} {\bibfnamefont {Q.}~\bibnamefont {Yu}}, \bibinfo {author} {\bibfnamefont {S.}~\bibnamefont {Yu}},\ and\ \bibinfo {author} {\bibfnamefont {H.}~\bibnamefont {Wu}},\ }\bibfield  {title} {\bibinfo {title} {{Observation of Dynamical Super-Efimovian Expansion in a Unitary Fermi Gas}},\ }\href {https://doi.org/10.1103/physrevlett.120.125301} {\bibfield  {journal} {\bibinfo  {journal} {Phys. Rev. Lett.}\ }\textbf {\bibinfo {volume} {120}},\ \bibinfo {pages} {125301} (\bibinfo {year} {2018})}\BibitemShut {NoStop}%
\bibitem [{\citenamefont {Makotyn}\ \emph {et~al.}(2014)\citenamefont {Makotyn}, \citenamefont {Klauss}, \citenamefont {Goldberger}, \citenamefont {Cornell},\ and\ \citenamefont {Jin}}]{Makotyn2014}%
  \BibitemOpen
  \bibfield  {author} {\bibinfo {author} {\bibfnamefont {P.}~\bibnamefont {Makotyn}}, \bibinfo {author} {\bibfnamefont {C.~E.}\ \bibnamefont {Klauss}}, \bibinfo {author} {\bibfnamefont {D.~L.}\ \bibnamefont {Goldberger}}, \bibinfo {author} {\bibfnamefont {E.~A.}\ \bibnamefont {Cornell}},\ and\ \bibinfo {author} {\bibfnamefont {D.~S.}\ \bibnamefont {Jin}},\ }\bibfield  {title} {\bibinfo {title} {{Universal dynamics of a degenerate unitary Bose gas}},\ }\href {https://doi.org/10.1038/nphys2850} {\bibfield  {journal} {\bibinfo  {journal} {Nat. Phys.}\ }\textbf {\bibinfo {volume} {10}},\ \bibinfo {pages} {116} (\bibinfo {year} {2014})}\BibitemShut {NoStop}%
\bibitem [{\citenamefont {Eigen}\ \emph {et~al.}(2017)\citenamefont {Eigen}, \citenamefont {Glidden}, \citenamefont {Lopes}, \citenamefont {Navon}, \citenamefont {Hadzibabic},\ and\ \citenamefont {Smith}}]{Eigen2017}%
  \BibitemOpen
  \bibfield  {author} {\bibinfo {author} {\bibfnamefont {C.}~\bibnamefont {Eigen}}, \bibinfo {author} {\bibfnamefont {J.~A.~P.}\ \bibnamefont {Glidden}}, \bibinfo {author} {\bibfnamefont {R.}~\bibnamefont {Lopes}}, \bibinfo {author} {\bibfnamefont {N.}~\bibnamefont {Navon}}, \bibinfo {author} {\bibfnamefont {Z.}~\bibnamefont {Hadzibabic}},\ and\ \bibinfo {author} {\bibfnamefont {R.~P.}\ \bibnamefont {Smith}},\ }\bibfield  {title} {\bibinfo {title} {{Universal Scaling Laws in the Dynamics of a Homogeneous Unitary Bose Gas}},\ }\href {https://doi.org/10.1103/physrevlett.119.250404} {\bibfield  {journal} {\bibinfo  {journal} {Phys. Rev. Lett.}\ }\textbf {\bibinfo {volume} {119}},\ \bibinfo {pages} {250404} (\bibinfo {year} {2017})}\BibitemShut {NoStop}%
\bibitem [{\citenamefont {Eigen}\ \emph {et~al.}(2018)\citenamefont {Eigen}, \citenamefont {Glidden}, \citenamefont {Lopes}, \citenamefont {Cornell}, \citenamefont {Smith},\ and\ \citenamefont {Hadzibabic}}]{Eigen2018}%
  \BibitemOpen
  \bibfield  {author} {\bibinfo {author} {\bibfnamefont {C.}~\bibnamefont {Eigen}}, \bibinfo {author} {\bibfnamefont {J.~A.~P.}\ \bibnamefont {Glidden}}, \bibinfo {author} {\bibfnamefont {R.}~\bibnamefont {Lopes}}, \bibinfo {author} {\bibfnamefont {E.~A.}\ \bibnamefont {Cornell}}, \bibinfo {author} {\bibfnamefont {R.~P.}\ \bibnamefont {Smith}},\ and\ \bibinfo {author} {\bibfnamefont {Z.}~\bibnamefont {Hadzibabic}},\ }\bibfield  {title} {\bibinfo {title} {{Universal prethermal dynamics of Bose gases quenched to unitarity}},\ }\href {https://doi.org/10.1038/s41586-018-0674-1} {\bibfield  {journal} {\bibinfo  {journal} {Nature}\ }\textbf {\bibinfo {volume} {563}},\ \bibinfo {pages} {221} (\bibinfo {year} {2018})}\BibitemShut {NoStop}%
\bibitem [{\citenamefont {Sykes}\ \emph {et~al.}(2014)\citenamefont {Sykes}, \citenamefont {Corson}, \citenamefont {D'Incao}, \citenamefont {Koller}, \citenamefont {Greene}, \citenamefont {Rey}, \citenamefont {Hazzard},\ and\ \citenamefont {Bohn}}]{Sykes2014}%
  \BibitemOpen
  \bibfield  {author} {\bibinfo {author} {\bibfnamefont {A.~G.}\ \bibnamefont {Sykes}}, \bibinfo {author} {\bibfnamefont {J.~P.}\ \bibnamefont {Corson}}, \bibinfo {author} {\bibfnamefont {J.~P.}\ \bibnamefont {D'Incao}}, \bibinfo {author} {\bibfnamefont {A.~P.}\ \bibnamefont {Koller}}, \bibinfo {author} {\bibfnamefont {C.~H.}\ \bibnamefont {Greene}}, \bibinfo {author} {\bibfnamefont {A.~M.}\ \bibnamefont {Rey}}, \bibinfo {author} {\bibfnamefont {K.~R.~A.}\ \bibnamefont {Hazzard}},\ and\ \bibinfo {author} {\bibfnamefont {J.~L.}\ \bibnamefont {Bohn}},\ }\bibfield  {title} {\bibinfo {title} {{Quenching to unitarity: Quantum dynamics in a three-dimensional Bose gas}},\ }\href {https://doi.org/10.1103/physreva.89.021601} {\bibfield  {journal} {\bibinfo  {journal} {Phys. Rev. A}\ }\textbf {\bibinfo {volume} {89}},\ \bibinfo {pages} {021601} (\bibinfo {year} {2014})}\BibitemShut {NoStop}%
\bibitem [{\citenamefont {Corson}\ and\ \citenamefont {Bohn}(2015)}]{Corson2015}%
  \BibitemOpen
  \bibfield  {author} {\bibinfo {author} {\bibfnamefont {J.~P.}\ \bibnamefont {Corson}}\ and\ \bibinfo {author} {\bibfnamefont {J.~L.}\ \bibnamefont {Bohn}},\ }\bibfield  {title} {\bibinfo {title} {{Bound-state signatures in quenched Bose-Einstein condensates}},\ }\href {https://doi.org/10.1103/physreva.91.013616} {\bibfield  {journal} {\bibinfo  {journal} {Phys. Rev. A}\ }\textbf {\bibinfo {volume} {91}},\ \bibinfo {pages} {013616} (\bibinfo {year} {2015})}\BibitemShut {NoStop}%
\bibitem [{\citenamefont {Corson}\ and\ \citenamefont {Bohn}(2016)}]{Corson2016}%
  \BibitemOpen
  \bibfield  {author} {\bibinfo {author} {\bibfnamefont {J.~P.}\ \bibnamefont {Corson}}\ and\ \bibinfo {author} {\bibfnamefont {J.~L.}\ \bibnamefont {Bohn}},\ }\bibfield  {title} {\bibinfo {title} {{Ballistic quench-induced correlation waves in ultracold gases}},\ }\href {https://doi.org/10.1103/physreva.94.023604} {\bibfield  {journal} {\bibinfo  {journal} {Phys. Rev. A}\ }\textbf {\bibinfo {volume} {94}},\ \bibinfo {pages} {023604} (\bibinfo {year} {2016})}\BibitemShut {NoStop}%
\bibitem [{\citenamefont {Colussi}\ \emph {et~al.}(2018)\citenamefont {Colussi}, \citenamefont {Corson},\ and\ \citenamefont {D'Incao}}]{Colussi2018}%
  \BibitemOpen
  \bibfield  {author} {\bibinfo {author} {\bibfnamefont {V.~E.}\ \bibnamefont {Colussi}}, \bibinfo {author} {\bibfnamefont {J.~P.}\ \bibnamefont {Corson}},\ and\ \bibinfo {author} {\bibfnamefont {J.~P.}\ \bibnamefont {D'Incao}},\ }\bibfield  {title} {\bibinfo {title} {{Dynamics of Three-Body Correlations in Quenched Unitary Bose Gases}},\ }\href {https://doi.org/10.1103/physrevlett.120.100401} {\bibfield  {journal} {\bibinfo  {journal} {Phys. Rev. Lett.}\ }\textbf {\bibinfo {volume} {120}},\ \bibinfo {pages} {100401} (\bibinfo {year} {2018})}\BibitemShut {NoStop}%
\bibitem [{\citenamefont {Sun}\ \emph {et~al.}(2020)\citenamefont {Sun}, \citenamefont {Zhang},\ and\ \citenamefont {Zhai}}]{Sun2020}%
  \BibitemOpen
  \bibfield  {author} {\bibinfo {author} {\bibfnamefont {M.}~\bibnamefont {Sun}}, \bibinfo {author} {\bibfnamefont {P.}~\bibnamefont {Zhang}},\ and\ \bibinfo {author} {\bibfnamefont {H.}~\bibnamefont {Zhai}},\ }\bibfield  {title} {\bibinfo {title} {{High Temperature Virial Expansion to Universal Quench Dynamics}},\ }\href {https://doi.org/10.1103/physrevlett.125.110404} {\bibfield  {journal} {\bibinfo  {journal} {Phys. Rev. Lett.}\ }\textbf {\bibinfo {volume} {125}},\ \bibinfo {pages} {110404} (\bibinfo {year} {2020})}\BibitemShut {NoStop}%
\bibitem [{\citenamefont {Huo}\ and\ \citenamefont {Zhang}(2026)}]{huo2026}%
  \BibitemOpen
  \bibfield  {author} {\bibinfo {author} {\bibfnamefont {Y.}~\bibnamefont {Huo}}\ and\ \bibinfo {author} {\bibfnamefont {P.}~\bibnamefont {Zhang}},\ }\href {https://arxiv.org/abs/2610.16126} {\bibinfo {title} {Few-magnon universality beyond contact interactions in long-range spin chains}} (\bibinfo {year} {2026}),\ \Eprint {https://arxiv.org/abs/2610.16126} {arXiv:2610.16126 [cond-mat.quant-gas]} \BibitemShut {NoStop}%
\bibitem [{SM()}]{SM}%
  \BibitemOpen
  \href@noop {} {}\bibinfo {note} {See Supplemental Material at [URL will be inserted by publisher] for the numerical settings for finite open chain simulation, and details of experimental protocol.}\BibitemShut {Stop}%
\bibitem [{\citenamefont {Wall}\ \emph {et~al.}(2017)\citenamefont {Wall}, \citenamefont {Safavi-Naini},\ and\ \citenamefont {Rey}}]{Wall2017}%
  \BibitemOpen
  \bibfield  {author} {\bibinfo {author} {\bibfnamefont {M.~L.}\ \bibnamefont {Wall}}, \bibinfo {author} {\bibfnamefont {A.}~\bibnamefont {Safavi-Naini}},\ and\ \bibinfo {author} {\bibfnamefont {A.~M.}\ \bibnamefont {Rey}},\ }\bibfield  {title} {\bibinfo {title} {Boson-mediated quantum spin simulators in transverse fields: {$XY$} model and spin--boson entanglement},\ }\href {https://doi.org/10.1103/PhysRevA.95.013602} {\bibfield  {journal} {\bibinfo  {journal} {Phys. Rev. A}\ }\textbf {\bibinfo {volume} {95}},\ \bibinfo {pages} {013602} (\bibinfo {year} {2017})}\BibitemShut {NoStop}%
\bibitem [{\citenamefont {Figgatt}\ \emph {et~al.}(2019)\citenamefont {Figgatt}, \citenamefont {Ostrander}, \citenamefont {Linke}, \citenamefont {Landsman}, \citenamefont {Zhu}, \citenamefont {Maslov},\ and\ \citenamefont {Monroe}}]{Figgatt2019}%
  \BibitemOpen
  \bibfield  {author} {\bibinfo {author} {\bibfnamefont {C.}~\bibnamefont {Figgatt}}, \bibinfo {author} {\bibfnamefont {A.}~\bibnamefont {Ostrander}}, \bibinfo {author} {\bibfnamefont {N.~M.}\ \bibnamefont {Linke}}, \bibinfo {author} {\bibfnamefont {K.~A.}\ \bibnamefont {Landsman}}, \bibinfo {author} {\bibfnamefont {D.}~\bibnamefont {Zhu}}, \bibinfo {author} {\bibfnamefont {D.}~\bibnamefont {Maslov}},\ and\ \bibinfo {author} {\bibfnamefont {C.}~\bibnamefont {Monroe}},\ }\bibfield  {title} {\bibinfo {title} {Parallel entangling operations on a universal ion-trap quantum computer},\ }\href {https://doi.org/10.1038/s41586-019-1427-5} {\bibfield  {journal} {\bibinfo  {journal} {Nature}\ }\textbf {\bibinfo {volume} {572}},\ \bibinfo {pages} {368} (\bibinfo {year} {2019})}\BibitemShut {NoStop}%
\bibitem [{\citenamefont {Zhu}\ \emph {et~al.}(2023)\citenamefont {Zhu}, \citenamefont {Green}, \citenamefont {Nguyen}, \citenamefont {Alderete}, \citenamefont {Mossman},\ and\ \citenamefont {Linke}}]{Zhu2023Parallel}%
  \BibitemOpen
  \bibfield  {author} {\bibinfo {author} {\bibfnamefont {Y.}~\bibnamefont {Zhu}}, \bibinfo {author} {\bibfnamefont {A.~M.}\ \bibnamefont {Green}}, \bibinfo {author} {\bibfnamefont {N.~H.}\ \bibnamefont {Nguyen}}, \bibinfo {author} {\bibfnamefont {C.~H.}\ \bibnamefont {Alderete}}, \bibinfo {author} {\bibfnamefont {E.}~\bibnamefont {Mossman}},\ and\ \bibinfo {author} {\bibfnamefont {N.~M.}\ \bibnamefont {Linke}},\ }\bibfield  {title} {\bibinfo {title} {Pairwise-parallel entangling gates on orthogonal modes in a trapped-ion chain},\ }\href {https://doi.org/10.1002/qute.202300056} {\bibfield  {journal} {\bibinfo  {journal} {Adv. Quantum Technol.}\ }\textbf {\bibinfo {volume} {6}},\ \bibinfo {pages} {2300056} (\bibinfo {year} {2023})}\BibitemShut {NoStop}%
\bibitem [{\citenamefont {Wang}\ and\ \citenamefont {Freericks}(2012)}]{Wang2012}%
  \BibitemOpen
  \bibfield  {author} {\bibinfo {author} {\bibfnamefont {C.-C.~J.}\ \bibnamefont {Wang}}\ and\ \bibinfo {author} {\bibfnamefont {J.~K.}\ \bibnamefont {Freericks}},\ }\bibfield  {title} {\bibinfo {title} {Intrinsic phonon effects on analog quantum simulators with ultracold trapped ions},\ }\href {https://doi.org/10.1103/PhysRevA.86.032329} {\bibfield  {journal} {\bibinfo  {journal} {Phys. Rev. A}\ }\textbf {\bibinfo {volume} {86}},\ \bibinfo {pages} {032329} (\bibinfo {year} {2012})}\BibitemShut {NoStop}%
\bibitem [{\citenamefont {Lewis}\ \emph {et~al.}(2023)\citenamefont {Lewis}, \citenamefont {Banchi}, \citenamefont {Teoh}, \citenamefont {Islam},\ and\ \citenamefont {Bose}}]{Lewis2023}%
  \BibitemOpen
  \bibfield  {author} {\bibinfo {author} {\bibfnamefont {D.}~\bibnamefont {Lewis}}, \bibinfo {author} {\bibfnamefont {L.}~\bibnamefont {Banchi}}, \bibinfo {author} {\bibfnamefont {Y.~H.}\ \bibnamefont {Teoh}}, \bibinfo {author} {\bibfnamefont {R.}~\bibnamefont {Islam}},\ and\ \bibinfo {author} {\bibfnamefont {S.}~\bibnamefont {Bose}},\ }\bibfield  {title} {\bibinfo {title} {Ion trap long-range {$XY$} model for quantum state transfer and optimal spatial search},\ }\href {https://doi.org/10.1088/2058-9565/acd953} {\bibfield  {journal} {\bibinfo  {journal} {Quantum Sci. Technol.}\ }\textbf {\bibinfo {volume} {8}},\ \bibinfo {pages} {035025} (\bibinfo {year} {2023})}\BibitemShut {NoStop}%
\end{thebibliography}%

\onecolumngrid          
\vspace{2\baselineskip} 
\begin{center}
  \textbf{\large End Matter}
\end{center}
\vspace{\baselineskip}
\twocolumngrid          

\appendix
\setcounter{secnumdepth}{1}
\section{Emergent Scaling Symmetry at Two-magnon Resonance}

We consider the homogeneous model in the $Q=0$ sector.
We first map the original spin model to a hard-core boson model, with $\sigma_i^-=b_i^\dagger$ and
$\sigma_i^z=1-2b_i^\dagger b_i$. After subtracting the vacuum energy and the uniform term
$4J_z\sum_i b_i^\dagger b_i$, the Hamiltonian becomes
\begin{equation}
\begin{aligned}
H&=-2\sum_{i<j}J_{ij}
(b_i^\dagger b_j+b_j^\dagger b_i)\\
&\quad-4J_z\sum_i
b_i^\dagger b_{i+1}^\dagger b_i b_{i+1}
+V\sum_i b_i^{\dagger 2}b_i^2,
\end{aligned}
\label{eq:s_scaling_bosonic_model}
\end{equation}
where $J_{ij}=J/|i-j|^\alpha$ and the limit $V\to+\infty$ imposes the hard-core constraint.
In the two-magnon sector, translational invariance allows us to write
$\Psi_Q(i,j;t)=\mathcal L^{-1/2} e^{\mathrm{i}Q(i+j)/2}\psi_Q(x,t)$
at fixed total quasimomentum $Q$, where $x=j-i\in\mathbb Z$ and $\mathcal L\to\infty$
is an auxiliary periodic-chain length.
Since the magnons are bosons, bosonic exchange symmetry also requires $\psi_Q(-x,t)=\psi_Q(x,t)$. In the relative-coordinate basis, the Hamiltonian is
\begin{equation}
\begin{aligned}
     H_Q=&-4J\sum_{x\in\mathbb Z}\sum_{\ell\ne0}
     \frac{\cos(Q\ell/2)}{|\ell|^\alpha}|x\rangle\langle x+\ell|
     +2V|0\rangle\langle0|\\&-4J_z\bigl(|1\rangle\langle1|+|-1\rangle\langle-1|\bigr).
     \label{eq:s_scaling_q0_hamiltonian}
\end{aligned}
\end{equation}

We now set $Q=0$ and write $\psi(x,t)=\psi_0(x,t)$. Via Fourier transformation of $\varphi(q,t)=\sum_x e^{-\mathrm{i}qx}\psi(x,t)$, 
we obtain the Schr\"odinger equation in the relative momentum space
\begin{equation}
\mathrm{i}\partial_t\varphi(q,t)=2\epsilon_q\varphi(q,t)+2V\psi(0,t)-8J_z\cos q\,\psi(1,t),
 \label{eq:s_scaling_sch}
\end{equation}
with $\epsilon_q=-4J\sum_{\ell=1}^\infty\cos(q\ell)/\ell^\alpha$ being the single-magnon dispersion.

To solve Eq.~\eqref{eq:s_scaling_sch} with the initial state
$\varphi_i(q)=\varphi(q,0)$, we transform to the energy domain,
$\tilde\varphi(q,\omega)=\int_0^\infty dt\,e^{\mathrm{i}\omega t}\varphi(q,t)$. Here we introduce $\omega=2\epsilon_0+E_+$, where $E_+=E+\mathrm{i}0^+$ and $E$ is real. Integration by parts gives $(\omega-2\epsilon_q)\tilde\varphi=\mathrm{i}\varphi_i+2V\tilde\psi(0)-8J_z\cos q\,\tilde\psi(1)$.
Using
$\tilde\psi(x,\omega)=\int_{-\pi}^{\pi}(dq/2\pi)
\cos(qx)\tilde\varphi(q,\omega)$ at $x=0,1$,
we eliminate $\tilde\psi(0,\omega)$ and
$\tilde\psi(1,\omega)$ to obtain
\begin{equation}
\begin{aligned}
\tilde\varphi(q,\omega)
&=\frac12\int_{-\pi}^{\pi}\frac{dq'}{2\pi}
\frac{\boldsymbol f(q)^{\mathsf T}G(\omega,0)
\boldsymbol f(q')\varphi_i(q')}
{(\omega-2\epsilon_q)(\omega-2\epsilon_{q'})}\\
&\quad+\frac{\mathrm{i}\varphi_i(q)}{\omega-2\epsilon_q},
\end{aligned}
\label{eq:s_scaling_exact}
\end{equation}
where $\boldsymbol f(q)=(1,\cos q)^{\mathsf T}$ and $G(\omega,0)$ is defined by
\begin{equation}
\begin{aligned}
\mathrm{i}\Sigma^{(ab)}(\omega,0)
&=\frac12\int_{-\pi}^{\pi}\frac{dq}{2\pi}
\frac{(\cos q)^{a+b}}{\omega-2\epsilon_q},\\
[-\mathrm{i}G(\omega,0)]^{-1}
&=\begin{pmatrix}
(4V)^{-1}&0\\
0&-(16J_z)^{-1}
\end{pmatrix}-\mathrm{i}\Sigma(\omega,0).
\end{aligned}
\label{eq:s_scaling_propagator}
\end{equation}
We recognize $G$ as the propagator of the auxiliary dimer fields for the on-site and nearest-neighbor pairs, with self-energy $\Sigma$, following the conventions of Ref.~\cite{Sun2026Efimov}.
The hard-core solution follows by setting $(4V)^{-1}=0$ in $G$, with an initial state satisfying $\psi(0,0)=0$.

At low momentum, $\epsilon_q-\epsilon_0\simeq u|q|^z$, where $z=\alpha-1\in(1/2,1)$ and $u=-4J\Gamma(1-\alpha)\sin(\pi\alpha/2)>0$. The self-energy has the threshold expansion~\cite{Sun2026Efimov}
\begin{equation}
\mathrm{i}\Sigma^{(ab)}(\omega,0)
=-\frac{C^{ab}}{J}+C_0u^{-1/z}(-E_+)^\nu+o(|E|^\nu),
\label{eq:s_scaling_lowenergy}
\end{equation}
where $\nu=(1-z)/z$,
$C_0=-2^{-1-1/z}/[z\sin(\pi/z)]>0$, and
$C^{ab}=-J\mathrm{i}\Sigma^{(ab)}(2\epsilon_0,0)$. Substituting this expansion into Eq.~\eqref{eq:s_scaling_propagator} and taking $V\to+\infty$ gives
\begin{widetext}
\begin{equation}
 G(\omega,0)\simeq
 \frac{\mathrm{i}\begin{pmatrix}
 \dfrac{C^{11}}{J}-\dfrac{1}{16J_z}-C_0u^{-1/z}(-E_+)^\nu
 &-\dfrac{C^{01}}{J}+C_0u^{-1/z}(-E_+)^\nu\\[4pt]
 -\dfrac{C^{01}}{J}+C_0u^{-1/z}(-E_+)^\nu
 &\dfrac{C^{00}}{J}-C_0u^{-1/z}(-E_+)^\nu
 \end{pmatrix}}
 {\dfrac{C^{00}C^{11}-(C^{01})^2}{J^2}
 -\dfrac{C^{00}}{16JJ_z}
 -C_0u^{-1/z}(-E_+)^\nu
 \left[\dfrac{C^{00}+C^{11}-2C^{01}}{J}-\dfrac{1}{16J_z}\right]}.
 \label{eq:s_scaling_G_lowenergy}
\end{equation}
\end{widetext}
At the two-magnon resonance, $\det\!\left[\bigl(-\mathrm{i}G(2\epsilon_0,0)\bigr)^{-1}\right]=0$. Using this condition and the low-energy approximation $\boldsymbol f(q)^{\mathsf T}\simeq(1,1)$, we obtain 
\begin{equation}
 \boldsymbol f(q)^{\mathsf T}G(\omega,0)\boldsymbol f(q')\simeq
 -\frac{\mathrm{i} u^{1/z}}{C_0(-E_+)^\nu}.
 \label{eq:s_scaling_resonance}
\end{equation}
For an initial state concentrated near $q=0$, substituting
this result into Eq.~\eqref{eq:s_scaling_exact} and extending
the momentum integral to the real axis gives the continuum
expression
\begin{equation}
\begin{aligned}
\tilde\varphi(q,\omega)
&\simeq-\frac{\mathrm{i}u^{1/z}(-E_+)^{-\nu}}
{2C_0(E_+-2u|q|^z)}
\int_{-\infty}^{\infty}\frac{dq'}{2\pi}
\frac{\varphi_i(q')}{E_+-2u|q'|^z}\\
&\quad +\frac{\mathrm{i}\varphi_i(q)}{E_+-2u|q|^z}.
\end{aligned}
\label{eq:s_scaling_universal}
\end{equation}

Transforming Eq.~\eqref{eq:s_scaling_universal} back to the time domain via
$\varphi(q,t)=e^{-2\mathrm{i}\epsilon_0t}\int_{-\infty}^{\infty}\frac{dE}{2\pi}
e^{-\mathrm{i}Et}\tilde\varphi(q,2\epsilon_0+E_+)$, we obtain
\begin{equation}
\begin{aligned}
\varphi(q,t)
&=\Biggl[-\mathrm{i}\int_0^t dt_1\,
e^{-2\mathrm{i}(\epsilon_0+u|q|^z)(t-t_1)}\\
&\quad\times\int_0^{t_1}dt_2\,h(t_1-t_2)F(t_2)\Biggr]
+e^{-2\mathrm{i}(\epsilon_0+u|q|^z)t}\varphi_i(q),
\end{aligned}
\label{eq:s_scaling_duhamel}
\end{equation}
where
\begin{align}
 F(t)&=\int_{-\infty}^{\infty}\frac{dq}{2\pi}e^{-2\mathrm{i}(\epsilon_0+u|q|^z)t}\varphi_i(q),\\
 h(t)&=-\frac{u^{1/z}e^{\mathrm{i}\pi\nu/2}}{2C_0\Gamma(\nu)}e^{-2\mathrm{i}\epsilon_0t}t^{\nu-1},\quad t>0.
 \label{eq:s_scaling_kernels}
\end{align}

Now we consider the space-time transformation $x\to \lambda x, t\to\lambda^zt$, together with the transformation of initial state
\begin{equation}
 \psi_{i}(x)\to\lambda^{-1/2}\psi_i(x/\lambda),\qquad
 \varphi_{i}(q)\to\lambda^{1/2}\varphi_i(\lambda q). 
 \label{eq:s_scaling_initial}
\end{equation}
Under these transformations, we obtain
\begin{align}
F(\lambda^zt)&=\lambda^{-1/2}e^{-2\mathrm{i}\epsilon_0(\lambda^z-1)t}F(t),\\
 h(\lambda^zt)&=\lambda^{z(\nu-1)}e^{-2\mathrm{i}\epsilon_0(\lambda^z-1)t}h(t).
 \label{eq:s_scaling_kernel_dilations}
\end{align}
We can evaluate Eq.~\eqref{eq:s_scaling_duhamel}
at $(q/\lambda,\lambda^zt)$ for the scaled initial state. The integration region becomes $0\le\tau_2\le\tau_1\le t$. The interaction term acquires the factor
$\lambda^{2z}\lambda^{z(\nu-1)}\lambda^{-1/2}=\lambda^{1/2}$ because $z(1+\nu)=1$, and the common phases combine into $e^{-2\mathrm{i}\epsilon_0(\lambda^z-1)t}$. The free term has the same factors, so
\begin{equation}
\varphi(q/\lambda,\lambda^zt)=\lambda^{1/2}e^{-2\mathrm{i}\epsilon_0(\lambda^z-1)t}\varphi(q,t).
\end{equation}
Finally, via inverse Fourier transformation, we obtain
\begin{equation}
    \psi(\lambda x,\lambda^zt)=\lambda^{-1/2}e^{-2\mathrm{i}\epsilon_0(\lambda^z-1)t}\psi(x,t),
    \label{eq:s_scaling_wavefunctions}
\end{equation}
which clearly shows the emergent scaling symmetry between evolutions from scaled initial states. 

\section{Two-magnon resonance of the microscopic lattice impurity model}
To analytically derive the microscopic resonance condition $J_z^*/J$ in the impurity case, we map the impurity model to a hard-core boson model,
\begin{equation}
\begin{aligned}
    H =& -2\sum_{i<j}J_{ij}(b_i^{\dag}b_j + b_j^{\dag}b_i) - 4J_zb_{m+1}^{\dag}b_{m}^{\dag}b_mb_{m+1} \\
    &+ V\sum_ib_i^{\dag}b_i^{\dag}b_ib_i. 
\end{aligned}
\end{equation}
The last term represents an on-site repulsive interaction where the $V\rightarrow\infty$ limit will be taken finally. 
Using the path-integral representation, it corresponds to the action
\begin{equation}
\begin{aligned}
S&=\int dt\sum_k\bar\psi_k(i\partial_t-\epsilon_k)\psi_k -\frac{\bar{d}_{m+1/2}^{(1)}d_{m+1/2}^{(1)}}{16J_z}\\
&-\frac{1}{2}\left(\bar{d}_{m+1/2}^{(1)}\psi_m\psi_{m+1} + \bar\psi_{m+1}\bar\psi_{m}d_{m+1/2}^{(1)}\right)  \\
&-\frac{1}{2}\sum_i\left(\bar{d}_i^{(0)}\psi_i\psi_i + \bar\psi_i\bar\psi_id_i^{(0)}\right) + \frac{1}{4V}\sum_i\bar{d}_i^{(0)}d_i^{(0)}, 
\end{aligned}
\end{equation}
with the auxiliary dimer fields $d^{(0)}$ and $d^{(1)}$ introduced via Hubbard-Stratonovich transformation. 
Here $\psi_i$ and $\bar\psi_i$ are bosonic coherent-state fields, and $\epsilon_k=-\sum_l4J\cos(kl)/l^\alpha$ is the single-magnon dispersion. 
The two-magnon scattering is described by the propagator of the auxiliary dimer fields, whose self-energy is written as
\begin{equation}
\hspace{-2ex}
i\Sigma^{(a,b)}(\omega,Q)=\frac{1}{2}\int_{-\pi}^\pi\frac{dq}{2\pi}\frac{\cos(q)^{a+b}}{\omega-\epsilon_{Q/2+q}-\epsilon_{Q/2-q}+i0^+},
\end{equation}
where $\omega=E+2\epsilon_0$ and $a,b\in\{0,1\}$.
Unlike the homogeneous model, here the impurity interaction acts only on the $m+1/2$ bond, whose auxiliary field $d^{(1)}$ couples to all momenta, while on-site auxiliary field $d^{(0)}$ remains diagonal in total momentum $Q$. By taking $V\rightarrow\infty$, we obtain the impurity propagator $D$ as
\begin{equation}
\begin{aligned}
&[-iD(\omega)]^{-1} = -\frac{1}{16J_z} \\
&\hspace{-2ex}
-\int_{-\pi}^{\pi}\frac{dQ}{2\pi}\Big[i\Sigma^{(11)}(\omega,Q) -\frac{i\Sigma^{(10)}(\omega,Q)i\Sigma^{(01)}(\omega,Q)}{i\Sigma^{(00)}(\omega,Q)}\Big]. 
\end{aligned}
\end{equation}
Solving the pole at $E=0$ by $[-iD(2\epsilon_0)]^{-1}=0$ gives the resonant point as
\begin{equation}
\dfrac{J_z^*}{J} = \left(16\int_{-\pi}^{\pi}\dfrac{dQ}{2\pi}\left[c_{11}(Q)-\dfrac{c_{10}(Q)^2}{c_{00}(Q)}\right]\right)^{-1}, 
\end{equation}
with
\begin{equation}
c_{ab}(Q)=\dfrac{J}{2}\int_{-\pi}^{\pi}\dfrac{dq}{2\pi}\dfrac{\cos(q)^{a+b}}{\epsilon_{\frac{Q}{2}+q}+\epsilon_{\frac{Q}{2}-q}-2\epsilon_0}. 
\end{equation}
Thus we obtain the microscopic lattice model parameter $J_z^*/J$ of the two-magnon resonance for the impurity model.

\end{document}


\title{Supplementary Material for 
``Emergent Dynamical Scale Invariance in Long-Range Quantum Spin Chains"}

\author{Xuhui Yao}
\affiliation{Department of Physics, Fudan University, Shanghai 200433, China}

\author{Ning Sun}
\email{ningsun.atom@gmail.com}
\affiliation{Department of Physics, Fudan University, Shanghai 200433, China}
\affiliation{Department of Physics, The University of Hong Kong, Hong Kong, China}

\author{Lei Feng}
\email{leifeng@fudan.edu.cn}
\affiliation{Department of Physics, Fudan University, Shanghai 200433, China}
\affiliation{Institute of Micronano Electronic Devices and Quantum Computing and Shanghai Key Laboratory of Meta-surface Optoelectronic Fine Control, Fudan University, Shanghai 200438, China}
\affiliation{Hefei National Laboratory, Hefei 230088, China}
\date{\today}
\maketitle

In this supplementary material, we present
(1) details of the efficient numerical simulations on finite open chains;
(2) details of the trapped-ion preparation and hybrid analog--digital
evolution protocol, together with numerical benchmarks.

\section{Fixed-chain simulations and finite-size scaling}
\label{sec:supp_full_space}
In this section, we present the numerical methods used to simulate
finite open chains, including exact time evolution in the
two-magnon Hilbert space and the rescaling protocol.
\subsection{Two-magnon Hamiltonians and numerical evolution}
We consider an open chain with sites $n=0,\ldots,L-1$ and write
the two-magnon state as
\begin{equation}
|\Psi\rangle=\sum_{0\le i<j<L}c_{ij}|i,j\rangle,
\qquad \sum_{i<j}|c_{ij}|^2=1.
\label{eq:s_ordered_two_magnon_state}
\end{equation}
The ordered basis counts each configuration once and has
dimension $L(L-1)/2$. In this basis, the Hamiltonian consists of single-magnon
hopping subject to the hard-core constraint and a diagonal
interaction term. We first introduce the one-magnon hopping matrix as
\begin{equation}
h_{ij}=\begin{cases}-2J/|i-j|^\alpha,&i\ne j,\\0,&i=j.\end{cases}
\label{eq:s_one_body_hopping}
\end{equation}
For the homogeneous model, subtracting the polarized-vacuum energy
from the Ising term gives the diagonal element of the Hamiltonian:
\begin{equation}
V^{\rm hom}_{ij}=8J_z-2J_z(\delta_{i,0}+\delta_{j,L-1})
-4J_z\delta_{j,i+1}.
\label{eq:s_homogeneous_diagonal}
\end{equation}
In particular, the boundary energies are retained. For the
impurity model, the interaction acts on the central bond:
\begin{equation}
V^{\rm imp}_{ij}=-4J_z\delta_{i,m}\delta_{j,m+1},
\qquad m=L/2-1,
\label{eq:s_impurity_diagonal}
\end{equation}
where all chain lengths used here are even.

To compute the time evolution $e^{-iHt}|\Psi(0)\rangle$,
we use \texttt{expm\_multiply} from SciPy 
\footnote{\url{https://docs.scipy.org/doc/scipy/reference/generated/scipy.sparse.linalg.expm_multiply.html}},
which requires repeated evaluations of $H|u\rangle$,
where $|u\rangle$ denotes an intermediate vector in the algorithm.
To evaluate these products efficiently, we expand
$|u\rangle=\sum_{i<j}u_{ij}|i,j\rangle$ and introduce
a symmetric matrix $\Phi_{ij}=\Phi_{ji}=u_{ij}/\sqrt{2}$
for $i<j$, with $\Phi_{ii}=0$.
The product in the ordered basis is then
\begin{equation}
(H|u\rangle)_{ij}=\sqrt{2}\,(h\Phi+\Phi h)_{ij}+V_{ij}u_{ij},
\qquad i<j,
\label{eq:product}
\end{equation}
with $V_{ij}=V^{\rm hom}_{ij}$ or $V^{\rm imp}_{ij}$. Restricting the output to
$i<j$ excludes double occupancy while retaining hops across the
other magnon to an unoccupied site. This formulation avoids explicitly storing the sparse
two-magnon Hamiltonian, reducing the storage requirement
from $\mathcal O(L^3)$ to $\mathcal O(L^2)$.

To reduce the computational cost, we further exploit the symmetric
Toeplitz structure of $h$ and evaluate $h\Phi$ using
fast Fourier transform (FFT)-based multiplication. 
Specifically, $h$ has the symmetric Toeplitz form $h_{ij}=g_{|i-j|}$, where $g_0=0$ and $g_r=-2J/r^\alpha$ for $1\le r<L$. Thus, $(h\Phi)_{ij}=\sum_{m=0}^{L-1}h_{im}\Phi_{mj}=\sum_{m=0}^{L-1}g_{|i-m|}\Phi_{mj}$. To evaluate this sum using FFTs, we rewrite it as a length-$K$ circular convolution,
\begin{equation}
y_i=\sum_{m=0}^{K-1}a_{(i-m)\bmod K}\phi^{(j)}_m.
\end{equation}
Here $\phi^{(j)}$ is column $j$ of $\Phi$ padded with zeros to length $K$. We choose $K\ge2L-1$ so that all displacements from $-(L-1)$ to $L-1$ have distinct indices modulo $K$. The kernel $a$ stores the corresponding hopping coefficients, with $a_0=0$, $a_r=a_{K-r}=g_r$ for $1\le r<L$, and all remaining entries zero. Consequently, $a_{(i-m)\bmod K}=h_{im}$ for $0\le i,m<L$, and the zero padding gives $y_i=(h\Phi)_{ij}$ for $0\le i<L$. By the convolution theorem, $y=\mathcal F_K^{-1}(\mathcal F_K[a]\odot\mathcal F_K[\phi^{(j)}])$, so we have
\begin{equation}
(h\Phi)_{ij}
=
\left[
\mathcal F_K^{-1}
\left(
\mathcal F_K[a]\odot\mathcal F_K[\phi^{(j)}]
\right)
\right]_i,
\qquad
0\le i<L,
\end{equation}
where $\mathcal F_K$ denotes the discrete Fourier transform
and $\odot$ denotes elementwise multiplication. 
With $K=\mathcal O(L)$, computing $h\Phi$ requires $L$
FFT-based column products, each costing $\mathcal O(L\log L)$. Since $h$ and $\Phi$ are symmetric,
$\Phi h=(h\Phi)^{\mathsf T}$.
Including the diagonal interaction and projection onto $i<j$,
the total cost per Hamiltonian--vector product is therefore
$\mathcal O(L^2\log L)$, compared with $\mathcal O(L^3)$
for direct sparse multiplication, while storage remains
$\mathcal O(L^2)$.

\subsection{Rescaling protocol in finite open chains}

To simulate low-energy scattering, we initialize a pair of
Gaussian magnon wave packets at each fixed chain length $L$:
\begin{equation}
\Psi(i,j;0)=\mathcal N_\lambda
\exp\left[-\frac{(i-c_{1,\lambda})^2+(j-c_{2,\lambda})^2}
{2\sigma_\lambda^2}+ik_\lambda(i-j)\right],
\qquad 0\le i<j<L,
\label{eq:s_full_packet}
\end{equation}
with $\lambda\in\{1,1.5,2,2.5,3\}$ and $\lambda=1$
denoting the smallest packet.
Here $\mathcal N_\lambda$ normalizes the state on the
finite ordered domain. Specifically, for both the homogeneous and impurity models,
we choose the two packet centers symmetrically
about the chain center and scale their separation, width, and
opposite momenta as
\begin{equation}
c_{1,\lambda}=(L-1)/2-\lambda d_0/2,\qquad
c_{2,\lambda}=(L-1)/2+\lambda d_0/2,\qquad
\sigma_\lambda=\lambda\sigma_0,\qquad
k_\lambda=k_0/\lambda,
\label{eq:s_full_packet_scaling}
\end{equation}
with $d_0=21$, $\sigma_0=5$, and $k_0=0.3$ at $L=180$.

For each evolved state, we compute the normalized one-body density
\begin{equation}
 \rho^{(1)}(j,t)=\frac12\langle\Psi(t)|P_\downarrow(j)|\Psi(t)\rangle
 =\frac12\left[\sum_{i<j}|\Psi(i,j;t)|^2
+\sum_{k>j}|\Psi(j,k;t)|^2\right].
\label{eq:s_one_body_density}
\end{equation}
The factor $1/2$ accounts for the two magnons so that $\sum_j\rho^{(1)}(j,t)=1$. The rescaling protocol for a finite open chain is implemented
by contracting spatial coordinates about the chain center
by a factor $\lambda$ and redistributing the probabilities
onto the original lattice bins.
Specifically, the rescaled probability used to calculate the scaling distance $\mathcal{D}(t)$ is given by
\begin{equation}
\widetilde{\rho}_\lambda^{(1)}(i,t)
=\sum_j w_{ij}^{(\lambda)}
\rho^{(1)}
(j,\lambda^z t), \qquad 0\le i\le L-1,
\label{eq:s_density_mapping}
\end{equation}
where the overlap weights are
\begin{equation}
w_{ij}^{(\lambda)}
=\lambda\left[
\min\left(i+1-\frac L2,\frac{j+1-L/2}{\lambda}\right)
-\max\left(i-\frac L2,\frac{j-L/2}{\lambda}\right)
\right]_+,
\end{equation}
with $[u]_+=\max(u,0)$. Note that for $\lambda\ge1$, $\sum_i w_{ij}^{(\lambda)}=1$, so $\sum_i\widetilde{\rho}_\lambda^{(1)}(i,t)=1$. The SD is then evaluated using the definition in the main
text, with the spatial integral replaced by a sum over
the lattice bins.
\begin{figure}[h]
    \centering
    \includegraphics[width=\linewidth]{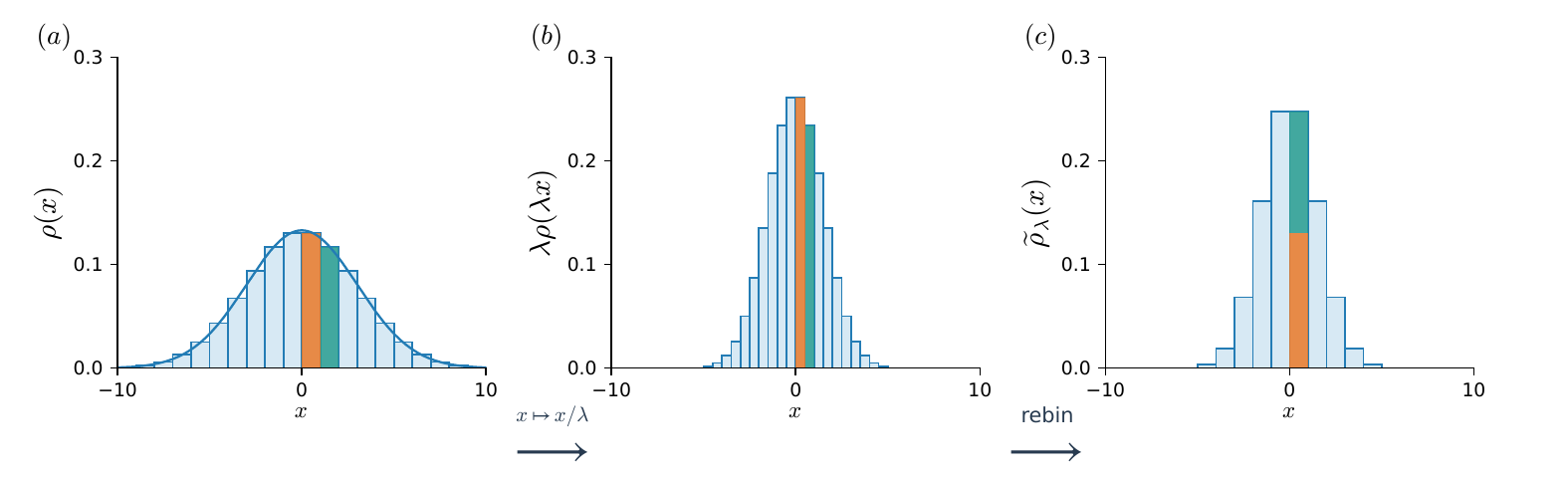}
\caption{\label{fig:rescaling_schematic}
    Schematic of the rescaling protocol on a finite open chain.
    (a) A Gaussian density on the original lattice.
    (b) Coordinate contraction $x\mapsto x/\lambda$.
    (c) Redistribution onto the common grid using the overlap weights
    $w_{ij}^{(\lambda)}$.
    Colored regions illustrate how two contracted bins contribute
    to one target bin while preserving their probabilities.
    }
\end{figure}
For the finite-size analysis, we repeat the scale comparison at
$L=180,240,300,360,420$ and
locate the minimum of $\overline{\mathcal D}$ near resonance.
When varying $L$, we set $\eta=L/180$ and use, for both models,
\begin{equation}
d_0=21\eta,\qquad \sigma_0=5\eta,\qquad k_0=0.3/\eta.
\label{eq:s_fss_packet_scaling}
\end{equation}
For the homogeneous (impurity) model, we sample $41$ ($61$)
uniformly spaced common times over $[0,T]$, including both
endpoints, with $JT=8(\eta/3)^{0.75}$ [$30(\eta/3)^{1.5}$].
The homogeneous scan covers $J_z/J\in[2.20,2.56]$ with spacing
$0.04$, refined to $0.005$ near the minima; the impurity scan
uses spacing $0.00125$ near the minima.
At each $L$, cubic-spline interpolation locates the SD minimum
within the sampled range. We fit these positions by unweighted
least squares to the finite-size ansatz in the main text.

\section{Experimental protocol and numerical benchmarks}
\label{sec:supp_experimental_protocol}

In this section, we describe the preparation and hybrid evolution underlying
Fig.~4 of the main text.
Our goal is to prepare the Gaussian two-magnon packets
defined in Eq.~\eqref{eq:s_full_packet},
with target centers $c_1^{\rm tar}$ and
$c_2^{\rm tar}$, width $\sigma$, and opposite
momenta $(+k,-k)$ (the scale index $\lambda$ is omitted for clarity).
Since the target centers may lie between lattice sites,
we start with localized spin flips at integer well centers
$c_1^{\rm ini}$ and $c_2^{\rm ini}$,
ensuring a unique initial ground state when the hopping is off.
We then broaden the wells and smoothly move their centers
to the target positions. Specifically, the preparation Hamiltonian is given by
\begin{equation}
H_{\rm prep}(t_{\rm prep})
=f(u)H_{XY}(\alpha_{\rm prep})
+\sum_n W_n(u)P_\downarrow(n),
\qquad u=\frac{t_{\rm prep}}{T_{\rm ramp}},
\quad 0\le u\le1.
\label{eq:s_prep_hamiltonian}
\end{equation}
Here $\alpha_{\rm prep}$ is the long-range hopping exponent during preparation.
The site-resolved longitudinal fields implement a symmetric
truncated-quadratic potential,
\begin{equation}
\begin{aligned}
W_n(u)
&=\min\!\left\{
\kappa(u)\min\!\left[
(n-c_1(u))^2,
(n-c_2(u))^2,R(u)^2
\right],\,W_{\max}\right\},\\
R(u)&=\frac{c_2(u)-c_1(u)-1}{2}.
\end{aligned}
\label{eq:s_moving_double_well}
\end{equation}
Here $R(u)$ is the truncation radius of the trap, $W_{\max}$ is the
upper bound on the potential energy, and $\kappa$ controls the trap curvature.
We turn on the hopping while reducing the confinement using
\begin{equation}
f(u)=10u^3-15u^4+6u^5,\qquad
\kappa(u)=\kappa_i+
(\kappa_f-\kappa_i)f(u).
\label{eq:s_prep_ramp}
\end{equation}
Here $\kappa_i$ and $\kappa_f$ are the initial and
final curvature coefficients. The centers remain fixed until
$u=u_c$, then move smoothly to their target positions:
\begin{equation}
\begin{aligned}
c_a(u)&=c_a^{\rm ini}
+q(u)(c_a^{\rm tar}-c_a^{\rm ini}),
\qquad a=1,2,\\
q(u)&=
\begin{cases}
0,&0\le u\le u_c,\\
f\!\left((u-u_c)/(1-u_c)\right),&u_c<u\le1,
\end{cases}
\end{aligned}
\label{eq:s_center_ramp}
\end{equation}
At $u=0$, the hopping vanishes and the unique two-magnon ground
state is $|c_1^{\rm ini},c_2^{\rm ini}\rangle$. The final curvature $\kappa_f$ is optimized separately
at each scale by maximizing the fidelity of the target state (with $k=0$) and the ground state of the preparation Hamiltonian. 
\begin{figure}[htbp]
    \centering
    \includegraphics[width=\linewidth]{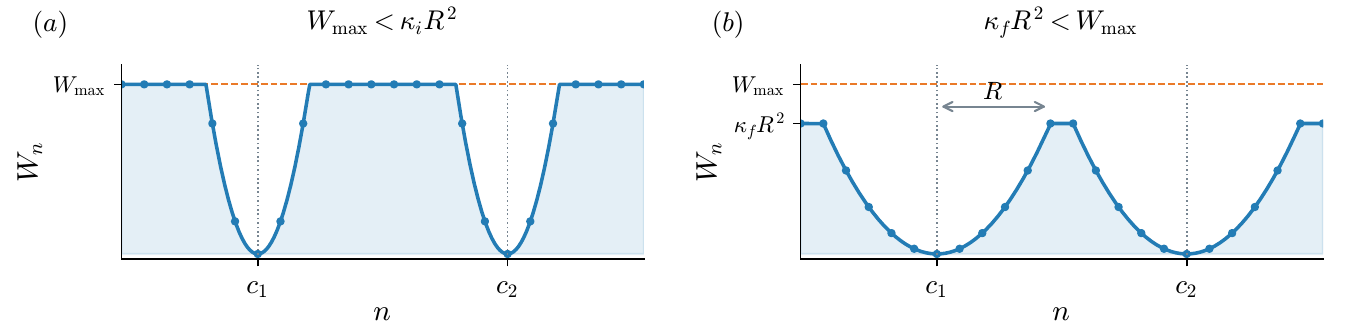}
    \caption{\label{fig:preparation_potential}
    Schematic of the symmetric truncated-quadratic potential.
    (a) Initially, strong confinement produces narrow wells
    capped at $W_{\max}$.
    (b) Reducing the curvature broadens the wells, with truncation
    set by the radius $R$ and plateau height $\kappa_f R^2$.
    Dots indicate lattice sites, and the dashed orange line
    marks $W_{\max}$.
    }
\end{figure}
After the adiabatic ramp, we imprint the site phase
via local $Z$ rotations:
\begin{equation}
\phi(n)=
\begin{cases}
k(n-c_1^{\rm tar}),&n\le(L-1)/2,\\
-k(n-c_2^{\rm tar}),&n>(L-1)/2.
\end{cases}
\label{eq:s_phase_imprint}
\end{equation}
Thus the ordered-coordinate amplitude at $(i,j)$ is multiplied
by $e^{i[\phi(i)+\phi(j)]}$.
The preparation fidelity $F_{\rm prep}$ defined in the main
text is evaluated after this phase imprinting, relative to
the moving Gaussian target.

For the preparation benchmark, we use an open chain of $L=180$
sites and $\lambda\in\{1,1.5,2,2.5,3\}$.
The target packets coincide with those used for the finite-chain
scattering calculations in Sec.~\ref{sec:supp_full_space}.
We choose $\alpha_{\rm prep}=3$, $\kappa_i=2J$,
$W_{\max}=12J$, and $u_c=0.2$.
Table~\ref{tab:s_preparation} lists the optimized curvatures,
ramp durations, and preparation fidelities.
All five scales satisfy $F_{\rm prep}>0.95$.

\begin{table}[ht]
\centering
\caption{Adiabatic preparation parameters at $L=180$.}
\label{tab:s_preparation}
\begin{ruledtabular}
\begin{tabular}{cccc}
$\lambda$ & $\kappa_f/J$ & $JT_{\rm ramp}$
& $F_{\rm prep}$\\
\hline
1   & $9.77\times10^{-3}$ & 25  & 0.9601\\
1.5 & $2.21\times10^{-3}$ & 90  & 0.9688\\
2   & $7.62\times10^{-4}$ & 170 & 0.9533\\
2.5 & $3.32\times10^{-4}$ & 320 & 0.9510\\
3   & $1.67\times10^{-4}$ & 550 & 0.9506\\
\end{tabular}
\end{ruledtabular}
\end{table}

Upon release, we remove the confining potential and evolve
with $H=H_{XY}+H_I$ using the symmetric second-order sequence
\begin{equation}
U(\Delta t)=
e^{-iH_{XY}\Delta t/2}
e^{-iH_I\Delta t}
e^{-iH_{XY}\Delta t/2}.
\label{eq:s_hybrid_trotter}
\end{equation}
Its single-step error is $\mathcal O(\Delta t^3)$, giving
a global evolution error of order $\Delta t^2$ at fixed total
time. For sufficiently small steps, the corresponding
state infidelity scales as $\Delta t^4$.

For the homogeneous interaction, the digital propagator is
\begin{equation}
e^{-iH_I^{\rm hom}\Delta t}
=\prod_{n=0}^{L-2}
e^{+iJ_z\Delta t\,\sigma_n^z\sigma_{n+1}^z}.
\label{eq:s_homogeneous_gate}
\end{equation}
All bond terms commute, but in experiment gates sharing an ion cannot be performed in parallel, so they are assigned to separate even- and odd-bond layers.
Each step therefore requires two entangling layers and
$L-1$ two-ion gates.
For the impurity interaction on the central bond
$m=L/2-1$, we obtain, up to a global phase,
\begin{equation}
e^{-iH_I^{\rm imp}\Delta t}
=
e^{-iJ_z\Delta t\,\sigma_m^z}
e^{-iJ_z\Delta t\,\sigma_{m+1}^z}
e^{+iJ_z\Delta t\,\sigma_m^z\sigma_{m+1}^z},
\label{eq:s_impurity_gate}
\end{equation}
Only one two-ion entangling gate and local $Z$ rotations
are required per step.
For the benchmark in Fig.~4(c) of the main text,
at $J\Delta t\simeq0.05$, the homogeneous implementation
requires $2N=158$--$360$ digital entangling layers
depending on $\lambda$.

\bibliographystyle{apsrev4-2}
\bibliography{ref}